\documentclass{article}
\usepackage{times}  % DO NOT CHANGE THIS
\usepackage{helvet}  % DO NOT CHANGE THIS
\usepackage{courier}  % DO NOT CHANGE THIS
\usepackage[hyphens]{url}  % DO NOT CHANGE THIS
\usepackage{graphicx} % DO NOT CHANGE THIS
\usepackage{caption} % DO NOT CHANGE THIS AND DO NOT ADD ANY OPTIONS TO IT
\usepackage{algorithm}
\usepackage{algorithmic}

\usepackage{times}  % DO NOT CHANGE THIS
\usepackage{helvet}  % DO NOT CHANGE THIS
\usepackage{courier}  % DO NOT CHANGE THIS
\usepackage[hyphens]{url}  % DO NOT CHANGE THIS
\usepackage{graphicx} % DO NOT CHANGE THIS
\usepackage{natbib}  % DO NOT CHANGE THIS AND DO NOT ADD ANY OPTIONS TO IT
\usepackage{caption} % DO NOT CHANGE THIS AND DO NOT ADD ANY OPTIONS TO IT
\DeclareCaptionStyle{ruled}{labelfont=normalfont,labelsep=colon,strut=off} % DO NOT CHANGE THIS
\usepackage{algorithm}
\usepackage{algorithmic}

\usepackage{xcolor}
\usepackage{newfloat}
\usepackage{listings}
\usepackage{newfloat}
\usepackage{listings}
\DeclareCaptionStyle{ruled}{labelfont=normalfont,labelsep=colon,strut=off} % DO NOT CHANGE THIS
\floatstyle{ruled}
\newfloat{listing}{tb}{lst}{}
\floatname{listing}{Listing}
\usepackage{authblk}

\usepackage[english]{babel}
\usepackage[letterpaper,top=2cm,bottom=2cm,left=3cm,right=3cm,marginparwidth=1.75cm]{geometry}

\usepackage{amsmath}
\usepackage{graphicx}
\usepackage[colorlinks=true, allcolors=blue]{hyperref}
\usepackage{csquotes}
\usepackage{subcaption}
\usepackage{wrapfig}
\usepackage{float}
\usepackage{longtable}
\usepackage{amssymb}
\usepackage{booktabs}
\usepackage{booktabs}
\usepackage{tikz}
\usetikzlibrary{positioning}
\usetikzlibrary{backgrounds}
\usetikzlibrary{arrows.meta}
\usetikzlibrary{fit}

\definecolor{posterPurpleDark}{HTML}{26215C}
\definecolor{posterPurpleMid}{HTML}{7F77DD}
\definecolor{posterPurpleLight}{HTML}{EEEDFE}
\definecolor{posterPurpleText}{HTML}{3C3489}
\definecolor{posterTealMid}{HTML}{1D9E75}
\definecolor{posterTealLight}{HTML}{E1F5EE}
\definecolor{posterTealText}{HTML}{0F6E56}
\definecolor{posterBlueMid}{HTML}{378ADD}
\definecolor{posterBlueLight}{HTML}{E6F1FB}
\definecolor{posterBlueText}{HTML}{0C447C}
\definecolor{posterCoralMid}{HTML}{D85A30}
\definecolor{posterCoralLight}{HTML}{FAECE7}
\definecolor{posterCoralText}{HTML}{712B13}
\definecolor{posterAmberLight}{HTML}{FAEEDA}
\definecolor{posterAmberText}{HTML}{633806}
\definecolor{posterGrayLight}{HTML}{E8E7E1}
\definecolor{posterGrayText}{HTML}{5F5E5A}
\definecolor{posterGrayRule}{HTML}{B4B2A9}
\definecolor{posterBg}{HTML}{F7F6F2}
\title{Who, Why, and How: Disentangling the Effects of Moderation Source, Context, and Language on Post-Removal Behavior}

\author{Siyi Zhou, Lindsay Young, Marlon Twyman, Emilio Ferarra}
\affil{ University of Southern California}

\begin{document}
\maketitle
\begin{abstract}
Content moderation is a central mechanism through which social media platforms 
attempt to balance user engagement with community governance. Yet existing research 
has largely treated moderation as a uniform intervention, overlooking how the source 
of moderation, the nature of the violation, and the linguistic style of removal 
explanations jointly shape user behavior. Drawing on the Human--AI Interaction 
Theory of Interactive Media Effects (HAII-TIME) framework, this study examines how 
these three dimensions interact to produce divergent post-moderation behavioral 
trajectories in a large-scale 
observational dataset of 11,795,036 moderation events across 9,285,410 users and 
61,261 subreddits on Reddit spanning January 2021 through December 2025. Using 
probabilistic behavioral classification, one-way ANOVA, and OLS regression with 
principal component analysis (PCA)-derived linguistic features, we find that 
automated bot moderation consistently produces higher compliance and lower 
self-censorship than both personal account and collective modteam moderation, 
challenging the assumption that human agency cues are inherently advantageous in 
human--AI interaction contexts. Modteam moderation produces the strongest 
self-censorship effects, suggesting that institutional depersonalization, rather 
than surveillance or enforcement uncertainty alone, is a meaningful driver of 
behavioral withdrawal. Violation severity emerges as a critical contingency: 
linguistic strategies that reduce resistance in routine moderation contexts, 
including elaborated explanation, community-scale numerical appeals, and direct 
personal address, can backfire for serious violations, whereas prosocially framed 
and emotionally emphatic messages become substantially more effective precisely 
when the stakes are highest. Of 480 linguistic interactions tested, 33 survive 
FDR correction, concentrated around harmful or illegal content and formatting 
violations. Together, these findings extend HAII-TIME by introducing violation 
salience as a moderator of cue-based processing, and offer empirical grounding 
for context-adaptive moderation design that can better balance platform governance 
with user engagement.

\noindent\textbf{Keywords:} content moderation, human--AI interaction, HAII-TIME, 
behavioral trajectories, self-censorship, reactance, compliance, linguistic features, 
Reddit, platform governance
\end{abstract}

\section{Introduction}

Social media platforms operate under competing and often contradictory logics: the pursuit of user engagement and the need to maintain a healthy online environment \cite{christin_internal_2024}. Content moderation serves as a central mechanism through which platforms attempt to reconcile these tensions by regulating the visibility and circulation of user-generated content. Moderation practices take multiple forms, including shadowbanning, deplatforming, and content removal \cite{gillespie_not_2022, myers_west_censored_2018, grimmelmann_virtues_2015}. Among these, content removal has received substantial scholarly attention, particularly regarding its effectiveness in promoting user compliance and sustaining engagement \cite{srinivasan_content_2019, jhaver_did_2019}.

However, existing research has largely treated content removal as a uniform intervention, overlooking the diversity of moderation practices that have emerged in contemporary platforms. In practice, moderation is carried out by a range of actors, including automated systems, individual human moderators, and collective moderator accounts, each operating under different incentives and constraints \cite{reddit_reddit_2026}. Moreover, the objectives of moderation vary widely, from preventing spam and enforcing formatting rules to addressing misinformation and harmful content \cite{jiang_characterizing_2020}. As a result, the effectiveness of who moderates and for what purpose remains underexplored. A more nuanced understanding of these distinctions is critical for identifying how platforms can better balance their competing logics of engagement and governance.

Content moderation is increasingly a collective effort between humans and automated systems, and thus should be understood as a form of human–computer interaction (HCI). Prior research suggests that transparency in moderation, such as providing explanations for content removal, can improve user engagement and perceptions of fairness \cite{jhaver_does_2019}. At the same time, the growing prevalence of social bots and AI systems has made automated moderation ubiquitous \cite{ferrara_rise_2016}. Consequently, users now receive moderation messages from both human and non-human agents, challenging existing theoretical frameworks in HCI.

Traditional perspectives offer only partial explanations for this dynamic. The Computers as Social Actors (CASA) framework conceptualizes computers as entities that elicit social responses from users \cite{nass_computers_1994}, whereas Computer-Mediated Communication (CMC) frameworks position computers as channels facilitating human–human interaction \cite{walther_computer-mediated_1996}. With the rise of AI and increasingly agentic systems, these distinctions are becoming blurred. Contemporary computational agents do not merely trigger social responses or mediate communication; they actively participate in communicative processes and shape interaction outcomes \cite{sundar_rise_2020}. In the context of content moderation, the coexistence of human and automated moderators reflects this shift toward hybrid communicative agency.

This study addresses two key gaps in the literature. First, prior research has largely overlooked the heterogeneity of moderation practices, treating content removal as a monolithic intervention despite variation in moderator identity, intent, and communication style. Second, there is limited theoretical understanding of how users respond to moderation in hybrid human–AI interaction contexts, where communicative agency is distributed across human and automated actors. By examining how different moderator roles, moderation purposes, and linguistic features jointly shape user behavior, this study provides a more nuanced account of moderation effectiveness.

Ultimately, this work aims to advance both theory and practice by identifying the conditions under which content moderation can effectively balance platform governance with user engagement.

\section{Literature Review}

\subsection{Content Moderation}

Social media platforms face a fundamental tension between enabling open participation 
and maintaining healthy discourse, making content moderation a necessary but 
consequential mechanism of platform governance \cite{baym_socially_2012, 
jenkins_convergence_2006, gillespie_custodians_2019}. Research consistently shows 
that moderation is not simply corrective: community-level interventions such as bans 
and quarantines often displace rather than eliminate harmful behavior 
\cite{chandrasekharan_you_2017, saleem_aftermath_2018, chandrasekharan_quarantined_2022, 
horta_ribeiro_platform_2021}, while at the individual level, moderation can generate 
chilling effects, reduce participation, encourage coded language, and in some cases 
intensify engagement with extreme content \cite{jhaver_evaluating_2021, 
 gerrard_beyond_2018, jhaver_bystanders_2024}. These outcomes 
suggest that how moderation is implemented matters as much as whether it is applied.

A key mechanism underlying these effects is users' perception of fairness, 
transparency, and control. Content removal raises longstanding concerns about 
platform governance legitimacy \cite{caplan_content_2018, gillespie_custodians_2019, 
roberts_censored_2018}, and users who feel their expression is constrained often 
experience reduced participation and diminished agency \cite{myers_west_censored_2018, 
devito_algorithms_2017}. How moderation is communicated plays a critical role here: 
directing users to generic community guidelines reduces perceived fairness, whereas 
transparent explanations improve trust \cite{goncalves_common_2023}. \citet{molina_when_2022} 
further demonstrate that trust in moderation depends on heuristic cues triggered by 
the perceived source, and that enabling user feedback enhances perceived agency. 
Although providing removal explanations can encourage continued engagement 
\cite{jhaver_does_2019}, it remains unclear whether this reflects genuine norm 
compliance or strategic circumvention. Together, these findings point to a critical 
gap: existing work remains focused on perceptual outcomes rather than on the 
direction and quality of post-moderation behavior.

\subsection{Human AI Interaction and Content Moderation}

The TIME framework conceptualizes two routes through which users interact with media: 
a cue route, whereby interface cues activate heuristic processing that shapes 
perceptions and judgments, and an action route, whereby feature use shapes users' 
sense of agency and behavioral engagement \cite{sundar_toward_2015}. HAII-TIME 
extends this to human and AI interaction contexts by emphasizing how perceived agency, 
whether attributed to human, artificial, or hybrid actors, shapes interpretation 
and evaluation of system outputs \cite{sundar_rise_2020}. Content moderation is a 
particularly apt context for this framework: moderation decisions attributed to 
human moderators, automated systems, or ambiguous collective accounts may lead users 
to form different interpretations of the legitimacy, flexibility, and negotiability 
of the intervention \cite{gillespie_custodians_2019, grimmelmann_virtues_2015}. 
While prior HAII-TIME applications have focused on perceptual outcomes such as trust 
\cite{molina_when_2022}, less is known about how these cue-based processes translate 
into behavioral responses, or how the communicative features of moderation messages 
shape behavioral adaptation over time \cite{jhaver_did_2019}.

To address this gap, we conceptualize post-moderation behavior through three 
theoretically grounded trajectories. The first, self-censorship, refers to a 
sustained reduction in posting activity following moderation, reflecting diminished 
user agency consistent with self-determination theory's prediction that reduced 
autonomy produces disengagement \cite{deci_what_2000}, and with chilling effects 
research showing that perceived enforcement leads individuals to suppress expression 
even without direct sanction \cite{schauer_fear_1978,penney_chilling_2016}. 
The second, compliance, refers to continued participation with behavioral adjustment, 
reflecting agency negotiation consistent with procedural justice theory, which holds 
that individuals comply more readily with rules they perceive as legitimate and 
fairly enforced \cite{tyler_why_1990}. The third, resistance, refers to continued 
rule violation following moderation, reflecting reactance against externally imposed 
constraints, consistent with psychological reactance theory's prediction that 
perceived threats to freedom motivate oppositional behavior 
\cite{brehm_theory_1966, dillard_nature_2005}. Linguistic features of moderation 
messages provide a further layer of influence, shaping how interventions are 
processed through heuristic cues or deeper elaboration \cite{petty_communication_1986, 
brown_politeness_1987, sundar_rise_2020}.

This framework motivates three research questions and one hypothesis. We predict 
that behavioral trajectories differ systematically across moderation sources:

\textbf{H1:} Users' post-moderation behavioral trajectories differ across sources 
of agency (e.g., bot, modteam, and personal account moderators).

Beyond source, violation type may signal varying levels of severity and 
interpretability that interact with source effects to shape behavior:

\textbf{RQ1:} How does the source of moderation interact with moderation context 
to shape users' post-moderation behavioral trajectories?

The linguistic style of removal explanations may further shape responses through 
heuristic or elaborative processing:

\textbf{RQ2:} How are linguistic features of moderation explanations associated 
with users' post-moderation behavioral trajectories?

\textbf{RQ3:} How do linguistic features moderate the relationship between 
moderation context, source of agency, and users' post-moderation behavioral 
trajectories?

\section{Methods}

\subsection{Data}

We use Reddit data spanning January 2021 through December 2025. Reddit is particularly well-suited for studying online moderation because its structure makes moderation actions and moderator identities directly observable, allowing us to link moderation events to subsequent user behavior at scale. Our data is obtained through academic torrent by \citet{raiderbdev_reddit_nodate}. Our platform-wide sample includes 11,795,036 moderation events across 9,285,410 users and 61,261 subreddits. Because Reddit moderators frequently leave removal explanation comments addressed to specific users, we are able to reconstruct individual-level behavioral trajectories before and after moderation, which is rare in observational platform data.

\subsection{Labeling}

\textbf{Moderator Classification.} We extract all comments tagged as moderator actions and classify them into three categories: (1) \textit{bot} accounts, identified by usernames containing ``bot'' or ``auto''; (2) \textit{modteam} accounts, identified by usernames containing ``modteam''; and (3) \textit{personal accounts}, which show no significant automated indicators and are presumed to be human-operated. This distinction matters because automated and team-based moderation may carry different social signals than moderation from a visible personal account, potentially producing different behavioral responses. We acknowledge that username-based classification is a heuristic proxy and may introduce some misclassification, particularly for bots using atypical naming conventions; however, this approach is consistent with prior computational work on moderator identity on Reddit \cite{jhaver_bystanders_2024, jhaver_does_2019}.

\textbf{Removal Explanations.} We extract all posts containing keywords indicative of content removal, excluding posts that resemble generic welcome messages for new members, as those do not constitute moderation actions directed at a specific user's behavior. All extracted posts were manually verified by a human annotator to confirm labeling accuracy. See Appendix for detailed explanation categorization procedure.

\textbf{Identifying Moderated Users.} We apply regular expression (regex) matching to identify instances where a removal explanation message references a username, specifically, patterns beginning with \texttt{/u} or \texttt{/u/} that are not embedded within a hyperlink. This allows us to directly link a moderation event to the specific user whose content was removed, which is necessary for tracking subsequent behavioral trajectories.

\textbf{Classifying Moderation Reasons.} We classify moderation reasons using keyword snowballing, iteratively expanding a seed set of keywords to capture the range of rule violations in the data. Keywords were then manually grouped into thematic categories by the research team. Capturing \textit{why} a user was moderated is essential for isolating whether different violation types produce systematically different behavioral responses, as theorized in RQ1.

\begin{figure*}[h!]
    \centering
    \begin{subfigure}[t]{0.32\textwidth}
        \includegraphics[width=\textwidth]{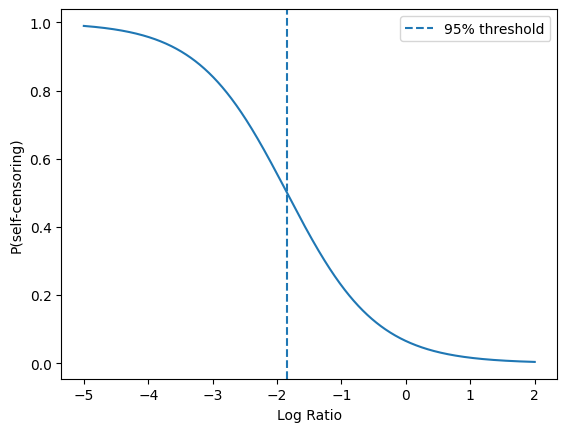}
        \caption{$P(Y=\text{self-censor} \mid \text{log ratio})$}
        \label{fig:selfcensor_classifier}
    \end{subfigure}
    \hfill
    \begin{subfigure}[t]{0.32\textwidth}
        \includegraphics[width=\textwidth]{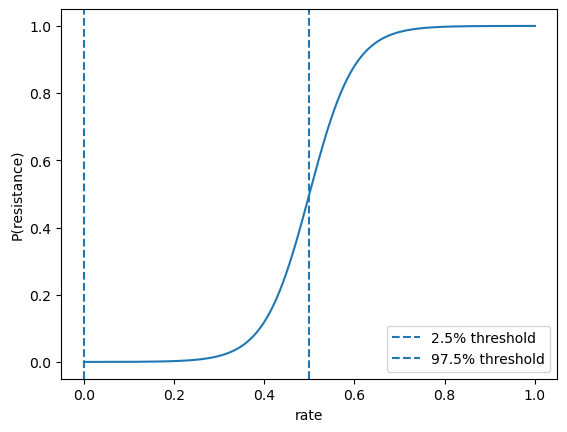}
        \caption{$P(Y=\text{resistant} \mid r)$}
        \label{fig:resistant}
    \end{subfigure}
    \hfill
    \begin{subfigure}[t]{0.32\textwidth}
        \includegraphics[width=\textwidth]{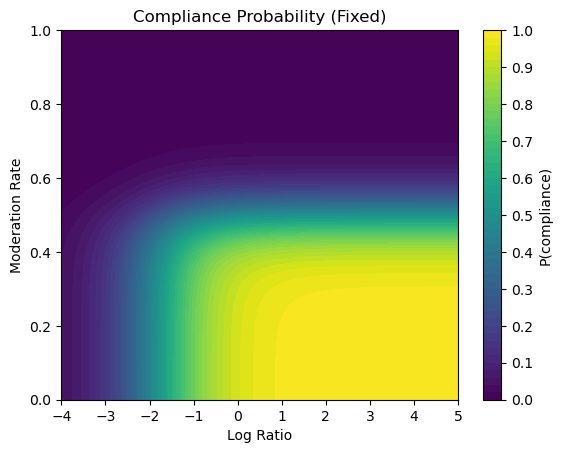}
        \caption{$P(Y=\text{compliance})$}
        \label{fig:compliance}
    \end{subfigure}
    \caption{Sigmoid-based probabilistic classifiers for self-censorship, resistance, and compliance.}
    \label{fig:classifiers}
\end{figure*}
\subsection{Measures}

\subsubsection*{Posting Frequency Change}

We define two complementary metrics to capture changes in user posting behavior before and after moderation. Using multiple metrics is intentional: each captures a different dimension of behavioral response, and relying on a single measure risks either privileging highly active users or obscuring meaningful variation among less active ones. Throughout, $N_{\text{before}}$ and $N_{\text{after}}$ denote the number of posts in a fixed window before and after moderation, respectively. We add 1 to $N_{\text{before}}$ in all measures to account for the removed post itself as part of pre-moderation activity and to avoid division by zero for users with no prior posts in the window.

\textbf{Direct Difference} measures the absolute change in posting volume:
\[
\Delta_{\text{post}} = N_{\text{after}} - (N_{\text{before}} + 1)
\]
This captures the raw magnitude of behavioral response, but because it is not normalized, highly active users can dominate group-level estimates.

\textbf{Log Ratio} measures proportional change in posting relative to baseline:
\[
\text{Log Ratio} = \log\!\left(\frac{N_{\text{after}}}{N_{\text{before}} + 1}\right)
\]
The log transformation symmetrizes the metric around zero and reduces the leverage of users with very high baseline activity, making it our primary metric for operationalizing self-censorship (see below).

\subsubsection*{Moderation Rate}

We measure the rate of subsequent moderation actions a user receives after their first removal:
\[
r = \frac{N_{\text{additional moderations}}}{N_{\text{posts after first moderation}} + N_{\text{additional moderations}}}
\]
Rather than using raw counts, we normalize by total post volume to ensure comparability across users who post at different frequencies. A user moderated 3 times out of 6 posts reflects a qualitatively different behavioral pattern than one moderated 3 times out of 300.

See appendix for all detailed distribution and threshold decision making. 

\subsubsection*{Probabilistic Behavior Classification via Sigmoid Mapping}

Rather than applying hard thresholds to classify behavioral outcomes, we use a sigmoid function to map behavioral metrics into continuous probability scores. This approach acknowledges that behavioral categories such as self-censorship and resistance exist on a continuum rather than as discrete states, and avoids the sensitivity to threshold placement that binary classification entails. The sigmoid function is defined as:
\[
\text{sigmoid}(x) = \frac{1}{1 + e^{k(x - \text{threshold})}}
\]
where $k$ controls the steepness of the transition and $\text{threshold}$ defines the decision boundary. We parameterize the slope $k$ using the interquartile range (IQR) of the observed distribution:
\[
k = \frac{2}{\text{IQR}}
\]
Anchoring $k$ to the IQR ensures the sigmoid's transition region spans the central mass of the data, where behavioral differences are most ambiguous and gradual probability assignment is most appropriate. This also makes the classifier adaptive to each metric's empirical distribution rather than imposing an arbitrary fixed slope.

\subsubsection*{Measuring Self-Censorship}

We operationalize self-censorship as a sustained reduction in posting activity following moderation. We use log ratio as the input metric because its distributional properties, such as symmetry around zero and reduced skew, make population-level thresholds more stable and interpretable. The decision boundary is set at the empirical 5th percentile of the population's monthly log-ratio distribution, capturing users whose posting decline is more extreme than 95\% of the general user population. This conservative threshold reflects our theoretical commitment to distinguishing genuine behavioral withdrawal from ordinary fluctuations in posting frequency. The resulting classifier is:
\[
P(Y = \text{self-censor} \mid \text{log ratio}) = \frac{1}{1 + e^{1.443(x + 1.846)}}
\]

\subsubsection*{Measuring Resistance}

We operationalize resistance as a persistently elevated moderation rate following an initial moderation event. The decision boundary is set at the 95th percentile of the population's moderation rate distribution, so that only users whose re-moderation rate is unusually high relative to the broader user base are classified as likely resistant:
\[
P(Y = \text{resistant} \mid r) = \frac{1}{1 + e^{-0.2(r - 0.5)}}
\]
Because the empirical moderation rate distribution is heavily zero-inflated, with both the 25th and 75th percentiles equal to zero, reflecting that most users are not re-moderated. Thus, the standard IQR-based slope is undefined. We therefore replace the degenerate IQR with a small upper bound of $[0, 0.1]$ and set $k = -20$, producing a steep but numerically stable sigmoid that closely approximates the intended hard threshold.

\subsubsection*{Measuring Compliance}

We define compliance as the joint absence of self-censorship and resistance, reflecting users who neither disengage from the community nor persist in rule-violating behavior. Rather than treating compliance as an independent third category, we derive it from the existing classifiers to ensure logical consistency across all three behavioral outcomes:
\[
P(Y = \text{compliance}) = (1 - P(\text{self-censor})) \times (1 - P(\text{resistant}))
\]

\subsection{Analysis}

Our analytic strategy maps directly onto the study's hypotheses and research questions. To test H1, which predicts differences in post-moderation behavioral trajectories across moderator types, we use one-way ANOVA to compare behavioral outcomes across bot, modteam, and personal account conditions. To address RQ1, which examines the interaction between moderation source and moderation context, we extend the ANOVA framework by incorporating moderation reason as a grouping factor, allowing us to assess whether source effects are consistent across violation types or contingent on context. For RQ2, which examines the relationship between linguistic features of moderation explanations and behavioral trajectories, we use linear regression with LIWC-derived features as predictors. Because LIWC produces a high-dimensional feature space, we first filter out features with a median of zero to remove variables with insufficient variation for meaningful analysis. We then apply Principal Component Analysis (PCA) to reduce dimensionality and surface latent linguistic dimensions that capture the most variance in the data. The resulting principal components serve as predictors in linear regression models, allowing us to interpret the relationship between linguistic style and post-moderation behavior in a lower-dimensional, less collinear space. Finally, to address RQ3, which examines how linguistic features moderate the relationship between source, context, and behavioral trajectories, we introduce interaction terms between the PCA-derived linguistic components and the moderation source and context variables in the regression models.
\subsection{Results}

\begin{table*}[htb]
\centering
\caption{Pairwise differences in behavioral outcomes across moderator types. 
Arrows indicate the direction and approximate magnitude of the difference for the 
row comparison (first group relative to second group). More arrows indicate larger 
differences. All reported pairwise differences are statistically significant after 
post-hoc correction unless marked with $\sim$.}
\label{tab:arrow_moderator_comparison}
\begin{tabular}{lccc}
\toprule
\textbf{Comparison} & \textbf{Self-censor} & \textbf{Resistance} & \textbf{Compliance} \\
\midrule
Bots vs. Modteam           & $\downarrow\downarrow\downarrow^{***}$ & $\sim^{*}$ & $\uparrow\uparrow\uparrow^{***}$ \\
Bots vs. Personal Accounts & $\downarrow\downarrow^{***}$           & $\downarrow\downarrow^{***}$ & $\uparrow\uparrow^{***}$ \\
Modteam vs. Personal Accounts & $\uparrow^{***}$                    & $\downarrow\downarrow^{***}$ & $\sim^{*}$ \\
\bottomrule
\end{tabular}
\vspace{0.5em}
\begin{minipage}{0.92\linewidth}
\footnotesize
\textit{Note.} 
$\uparrow$ indicates the first group has a higher value than the second; 
$\downarrow$ indicates the first group has a lower value than the second; 
$\sim$ indicates a statistically significant but substantively negligible difference. 
Arrow magnitude is based on pairwise median and mean differences. 
For self-censorship, the ordering is Modteam $>$ Personal Accounts $>$ Bots. 
For compliance, Bots show the highest levels, while Modteam and Personal Accounts 
are similar in median terms. For resistance, all three groups have median values 
of 0, so significant pairwise differences likely reflect the very large sample 
size rather than substantial behavioral separation.
\end{minipage}
\end{table*}

Table~\ref{tab:arrow_moderator_comparison} presents pairwise differences in 
behavioral outcomes across moderator types, offering support for H1. Bot moderation 
produces the highest compliance and the lowest self-censorship relative to both 
human moderator types, suggesting that users interpret automated enforcement as 
impersonal and rule-bound rather than as a social judgment. The strongest 
self-censorship effect emerges among modteam-moderated users, who show 
substantially higher withdrawal rates than both bot-moderated users (mean 
difference $= 0.085$, $p < .0001$) and personal account-moderated users (mean 
difference $= 0.013$, $p < .0001$). Personal account moderation occupies an 
intermediate position (mean difference from bots $= 0.072$, $p < .0001$). For 
compliance, bots outperform modteam by $0.078$ and personal accounts by $0.045$ 
(both $p < .0001$), while modteam and personal accounts converge at similar levels, 
suggesting that the human-versus-automated distinction drives compliance differences 
more than variation within human moderator types. Resistance differences reach 
statistical significance across all comparisons but should be interpreted with 
caution, as all groups share a median resistance value of zero, and differences 
most plausibly reflect large-sample sensitivity to distributional shape.

\begin{figure*}[h!]
    \centering
    \includegraphics[width=\textwidth]{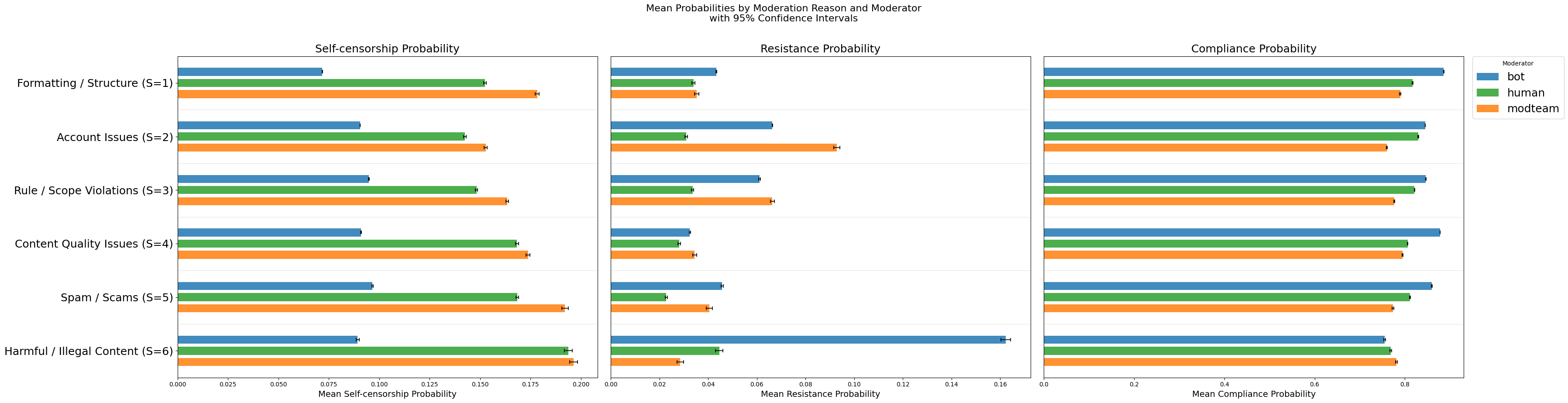}
    \caption{Mean probability of user behavior trajectory after moderation by 
    different source for different reasons. S denotes the severity level.}
    \label{fig:mean_behav_prob}
\end{figure*}

To address RQ1, Figure~\ref{fig:mean_behav_prob} presents mean behavioral 
probabilities across violation categories and moderator types. Kruskal-Wallis 
tests confirmed significant differences across moderation reasons for all three 
outcomes (self-censorship: $H = 217{,}212.25$, $p < .0001$; resistance: $H = 
29{,}907.54$, $p < .0001$; compliance: $H = 250{,}207.77$, $p < .0001$). A 
clear severity gradient emerges: harmful or illegal content produces the highest 
self-censorship ($M = 0.145$), the highest resistance ($M = 0.097$), and the 
lowest compliance ($M = 0.765$), while formatting or structure violations produce 
the lowest self-censorship ($M = 0.081$) and highest compliance ($M = 0.878$). 
The moderator source effect persists across all violation categories but widens 
with violation severity: the gap between bot and human moderator types is modest 
for formatting violations but substantially larger for harmful or illegal content. 
Resistance findings again warrant caution given near-zero median values 
($\tilde{x} = 0.000045$) across all categories.

\begin{figure*}[ht!]
    \centering
    \includegraphics[width=\textwidth]{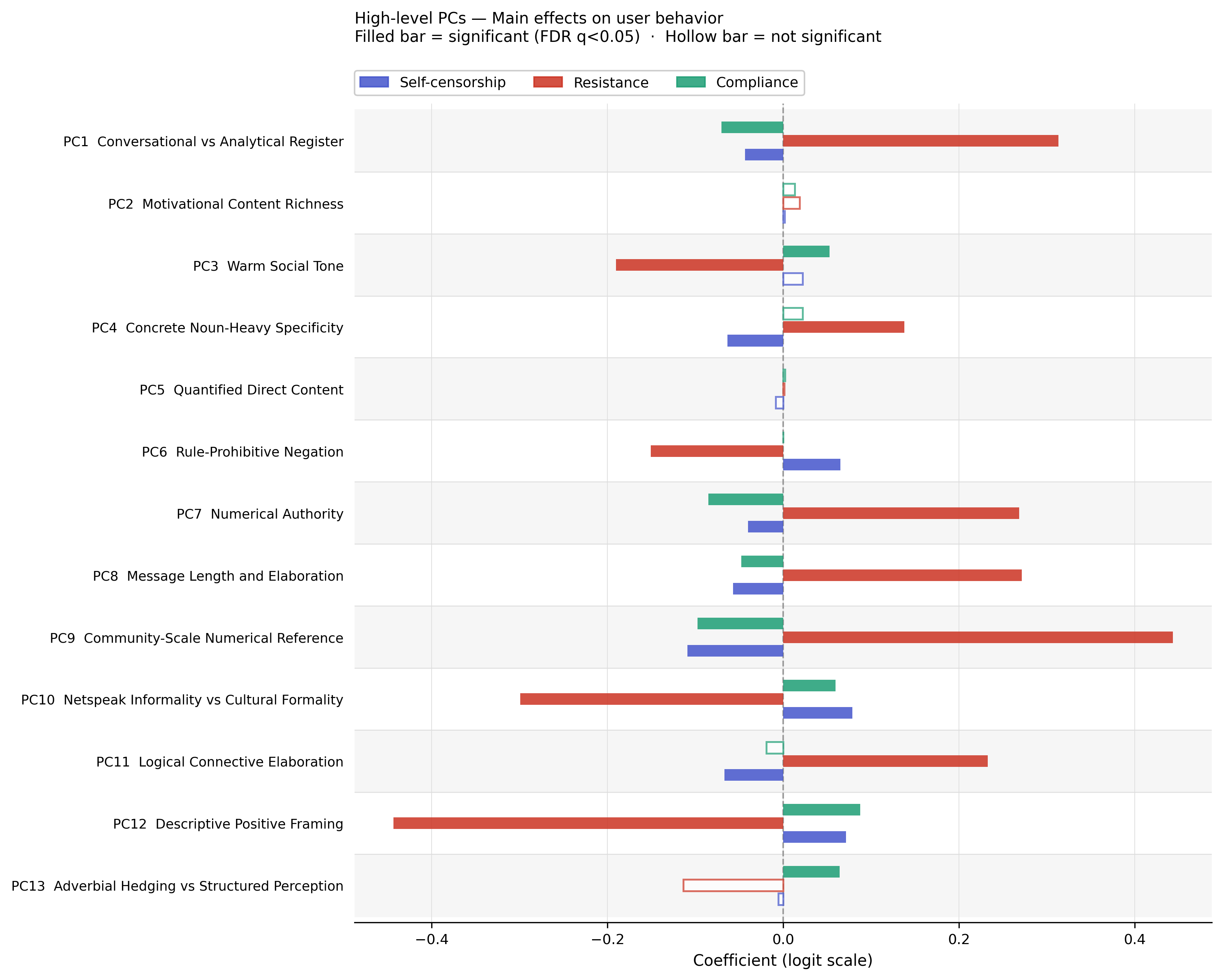}
    \caption{}
    \label{fig:main_HL}
\end{figure*}

\begin{figure*}[ht!]
    \centering
    \includegraphics[width=\textwidth]{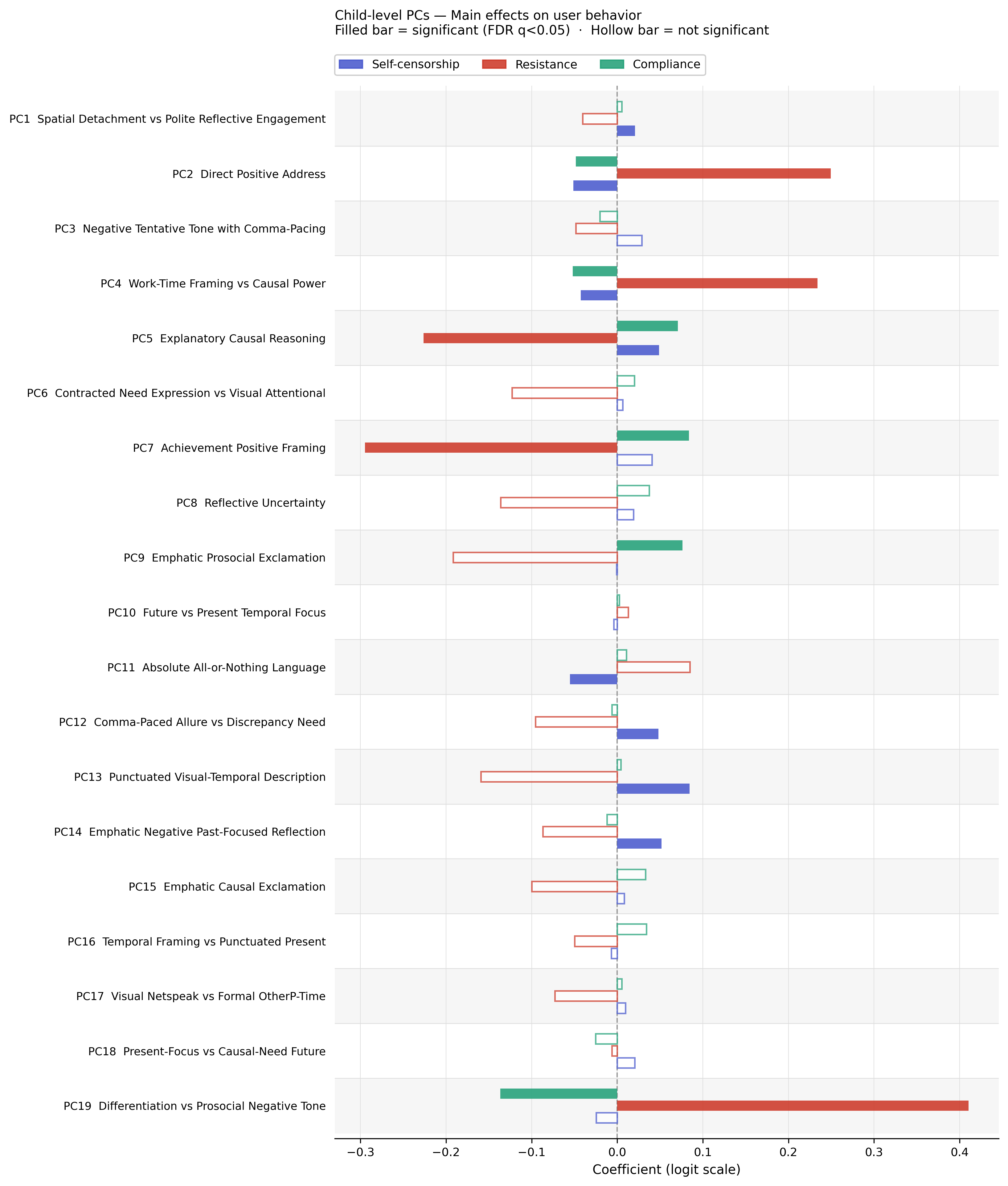}
    \caption{}
    \label{fig:main_CH}
\end{figure*}

To address RQ2, Figures~\ref{fig:main_HL} and~\ref{fig:main_CH} display the main 
effects of high-level (HL) and child-level (CH) principal components on behavioral 
outcomes, with filled bars indicating effects surviving FDR correction ($q < .05$). 
A consistent pattern emerges: formal, authoritative, or elaborated communication 
tends to increase resistance, while warm, prosocial, or positively framed 
communication tends to support compliance and reduce resistance. Among the HL PCs, 
Community-Scale Numerical Reference (HL-PC9; $\hat{\beta} = 0.443$, $q < .05$), 
Conversational vs.\ Analytical Register (HL-PC1; $\hat{\beta} = 0.313$, $q < .05$), 
and Message Length and Elaboration (HL-PC8; $\hat{\beta} = 0.271$, $q < .05$) show 
the strongest resistance-increasing effects, while Descriptive Positive Framing 
(HL-PC12; $\hat{\beta} = -0.443$, $q < .05$) and Warm Social Tone 
(HL-PC3; $\hat{\beta} = -0.190$, $q < .05$) are associated with lower resistance. 
Among the CH PCs, Explanatory Causal Reasoning (CH-PC5; $\hat{\beta} = -0.226$, 
$q < .05$) and Achievement Positive Framing (CH-PC7; $\hat{\beta} = -0.295$, 
$q < .05$) are the strongest resistance-reducing dimensions, while Direct Positive 
Address (CH-PC2; $\hat{\beta} = 0.249$, $q < .05$ for resistance; $\hat{\beta} = 
-0.051$, $q < .05$ for self-censorship) presents a nuanced pattern: warm direct 
engagement simultaneously reduces withdrawal while provoking resistance.

\begin{figure*}[ht!]
    \centering
    \includegraphics[width=\textwidth]{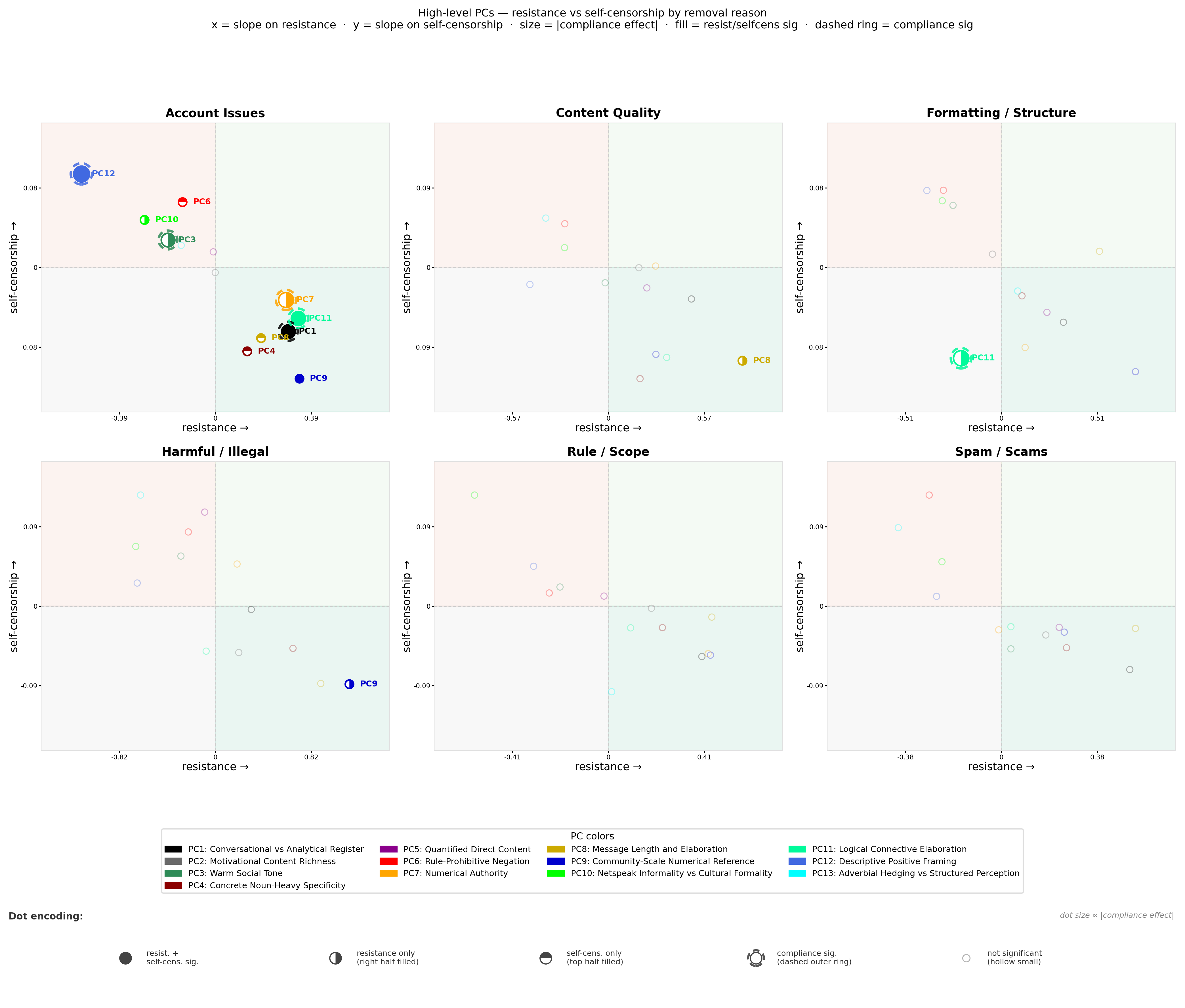}
    \caption{}
    \label{fig:interaction_HL}
\end{figure*}

\begin{figure*}[ht!]
    \centering
    \includegraphics[width=\textwidth]{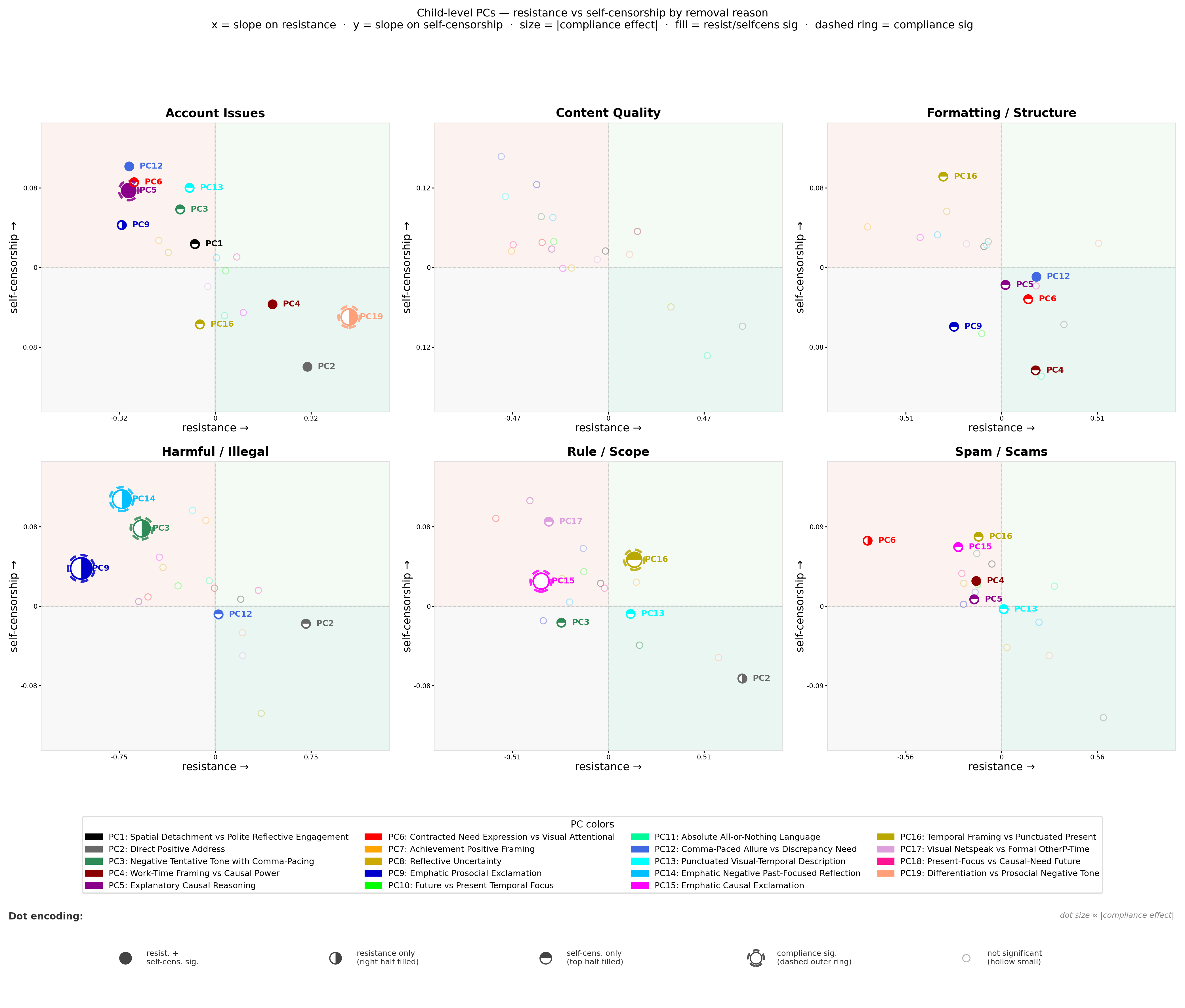}
    \caption{}
    \label{fig:interaction_CH}
\end{figure*}

To address RQ3, Figures~\ref{fig:interaction_HL} and~\ref{fig:interaction_CH} 
display effective slopes across violation categories. Of 480 interactions tested, 
only 33 survive FDR correction, a sparsity that is itself informative: linguistic 
effects are largely consistent across moderation contexts, with meaningful exceptions 
concentrated around harmful or illegal content and formatting violations. Notably, 
several features that show significant main effects do not produce significant 
interactions with context. Warm Social Tone, for instance, reduces resistance as a 
main effect but does not emerge as a significant moderator within any specific 
violation category, suggesting its benefit is diffuse rather than context-specific 
and may be diluted when violation severity introduces competing processing demands.

The distribution of significant interactions also reveals important asymmetry across 
violation types. Account issues attract the largest number of significant linguistic 
features that can be leveraged to improve compliance through removal explanation 
messages, offering moderators a relatively rich toolkit of communicative strategies. 
In contrast, harmful or illegal content has very few linguistic features that 
reliably improve outcomes. For this most severe violation category, the clearest 
interaction effects involve features associated with negative tone and past-focused 
reflection: Emphatic Negative Past-Focused Reflection (CH-PC14) amplifies 
self-censorship for harmful content ($\Delta = +0.10$, $q < .05$) but shows little 
effect on compliance and no significant effect on resistance. This asymmetry suggests 
that for the most serious violations, moderation messages can stop users from 
continuing to participate, but cannot readily redirect them toward compliant 
re-engagement. Harmful content violations, in other words, appear stoppable but not 
easily changeable through linguistic intervention alone.

Among the high-level PCs, Message Length and Elaboration (HL-PC8) dramatically 
amplifies resistance for content quality violations (effective slope $= 0.80$, 
$\Delta = 0.62$, $q < .01$) and harmful or illegal content (effective slope $= 
0.90$, $\Delta = 0.71$, $q < .05$) compared to $0.19$ for account issues, while 
Community-Scale Numerical Reference (HL-PC9) produces the largest resistance effect 
in the model for harmful content (effective slope $= 1.14$, $\Delta = 0.80$, $q < 
.05$). Logical Connective Elaboration (HL-PC11) reveals a sign reversal, increasing 
resistance for account issues but reducing it for formatting violations ($\Delta = 
-0.55$, $q < .05$), consistent with the interpretation that reasoned explanations 
are only effective when violations have clear, correctable rules. At the child level, 
Emphatic Prosocial Exclamation (CH-PC9) nearly triples its resistance-reduction 
effect for harmful content (effective slope $= -1.05$ vs. $-0.32$ for account 
issues, $\Delta = -0.73$, $q < .05$), while Direct Positive Address (CH-PC2) 
consistently amplifies resistance across serious violation categories, rising from 
$0.31$ for account issues to $0.71$ for both rule or scope violations and harmful 
content. Together, these findings confirm that the effectiveness of linguistic style 
is highly contingent on violation context, and that strategies effective in routine 
moderation can backfire for serious violations.
\section{Discussion}

This study examined how moderation source, violation context, and linguistic features 
jointly shape post-moderation behavioral trajectories on Reddit. Drawing on HAII-TIME 
\cite{sundar_rise_2020}, we theorized that moderation conveys agency cues producing 
divergent behavioral responses through psychological reactance 
\cite{brehm_theory_1966, dillard_nature_2005}, procedural justice 
\cite{tyler_why_1990}, and self-determination theory \cite{deci_what_2000}.

Bot moderation consistently produces higher compliance and lower self-censorship than 
both personal account and modteam moderation, challenging the assumption that human 
agency cues are inherently advantageous \cite{nass_computers_1994, sundar_rise_2020, 
molina_when_2022}. We interpret this as an \textit{impersonality advantage}: 
automated attribution signals rule-bound, non-judgmental enforcement, reducing the 
personal threat that drives reactance and autonomy loss \cite{brehm_theory_1966, 
deci_what_2000, binns_its_2018}. Modteam accounts produce the strongest 
self-censorship through what we term \textit{institutional depersonalization}: 
collective accounts are neither clearly human nor clearly automated, making appeal 
feel impossible and producing withdrawal from perceived absence of agency 
\cite{deci_what_2000, gillespie_custodians_2019, grimmelmann_virtues_2015}.

Linguistic cues do not operate uniformly across violation contexts but are moderated 
by violation severity, extending HAII-TIME \cite{sundar_rise_2020} with a 
\textit{violation salience moderator}. We observe an \textit{irony of effort}: 
elaborate messages amplify resistance for serious violations, suggesting that 
extensive justification inadvertently signals intervention gravity and triggers 
reactance \cite{brehm_theory_1966, dillard_nature_2005}, whereas reasoned 
explanations improve compliance only when violations have clear, correctable rules 
\cite{tyler_why_1990, jhaver_does_2019}. Several features with significant main 
effects, including Warm Social Tone, lose significance when violation context is 
accounted for, suggesting their benefits are diffuse rather than context-specific. 
More broadly, account issues offer moderators a rich linguistic toolkit for shaping 
behavior, while harmful or illegal content offers very few effective levers. For the 
most serious violations, moderation can suppress participation but cannot redirect 
users toward compliance: such violations are stoppable but not easily changeable 
through linguistic intervention alone. Platforms should therefore not expect message 
design to rehabilitate users who post harmful content, and should instead invest in 
account-level interventions and graduated enforcement pathways 
\cite{gillespie_custodians_2019, braithwaite_responsive_2001}. For lower-severity 
violations, context-adaptive messaging, short prosocial framing for serious cases 
and logically elaborated guidance for procedural ones, represents a tractable 
intervention \cite{jhaver_does_2019, chandrasekharan_internets_2018}.

\subsection{Limitations and Future Directions}

Several limitations warrant acknowledgment. First, our outcomes are behavioral 
rather than perceptual: we observe downstream traces of reactance and legitimacy 
judgments without measuring underlying psychological mechanisms. Future work 
combining observational data with experimental methods could address this. Second, 
the observational design precludes causal inference; we treat this as a 
quasi-natural experiment, and future experimental work would strengthen causal 
claims. Third, our moderator classification relies on username heuristics that may 
introduce measurement error, and Reddit's unusually transparent moderation structure 
may limit generalizability to platforms where moderation is less visible.

\subsection{Ethical Considerations}

This study carries potential risks that warrant acknowledgment. The findings could 
be misused by platform operators to engineer moderation messages that prioritize 
retention over genuine norm compliance or user wellbeing. We emphasize that our 
findings are intended to inform transparent, fairness-oriented moderation design 
rather than optimize behavioral outcomes as ends in themselves. The asymmetry we 
document between violation types could also be interpreted as justification for 
harsher enforcement of serious violations; we caution against this reading, as our 
findings speak to the limits of linguistic intervention rather than the appropriate 
scope of punitive action \cite{gillespie_custodians_2019, roberts_censored_2018}. 
Finally, findings may not generalize to platforms serving more vulnerable 
populations, where the consequences of moderation are likely more severe and where 
design implications should be applied with particular care.

\newpage
\bibliographystyle{aaai2026}
\bibliography{references}

@misc{reddit_reddit_2026,
	title = {Reddit {Rules}},
	url = {https://www.redditinc.com/policies/reddit-rules},
	publisher = {Reddit},
	author = {Reddit},
	year = {2026},
}

@book{tyler_why_1990,
	address = {New Haven},
	title = {Why people obey the law},
	isbn = {0-300-04403-8},
	publisher = {Yale University Press},
	author = {Tyler, Tom R.},
	year = {1990},
}

@book{roberts_censored_2018,
	address = {Princeton ;},
	title = {Censored : distraction and diversion inside {China}'s great firewall},
	isbn = {978-1-4008-9005-7},
	publisher = {Princeton University Press},
	author = {Roberts, Margaret E.},
	year = {2018},
}

@book{petty_communication_1986,
	address = {New York},
	series = {Springer series in social psychology},
	title = {Communication and persuasion : central and peripheral routes to attitude change},
	isbn = {0-387-96344-8},
	publisher = {Springer-Verlag},
	author = {Petty, Richard E. and Cacioppo, John T.},
	year = {1986},
}

@article{penney_chilling_2016,
	title = {Chilling {Effects}: {Online} {Surveillance} and {Wikipedia} {Use}},
	volume = {31},
	issn = {10863818, 23804742},
	url = {http://www.jstor.org/stable/43917620},
	number = {1},
	urldate = {2026-05-29},
	journal = {Berkeley Technology Law Journal},
	publisher = {[University of California, Berkeley, University of California, Berkeley, School of Law]},
	author = {Penney, Jonathon W.},
	year = {2016},
	pages = {117--182},
}

@inproceedings{jiang_characterizing_2020,
	address = {New York, NY, USA},
	series = {{CSCW} '20 {Companion}},
	title = {Characterizing {Community} {Guidelines} on {Social} {Media} {Platforms}},
	isbn = {978-1-4503-8059-1},
	url = {https://doi.org/10.1145/3406865.3418312},
	doi = {10.1145/3406865.3418312},
	booktitle = {Companion {Publication} of the 2020 {Conference} on {Computer} {Supported} {Cooperative} {Work} and {Social} {Computing}},
	publisher = {Association for Computing Machinery},
	author = {Jiang, Jialun 'Aaron' and Middler, Skyler and Brubaker, Jed R. and Fiesler, Casey},
	year = {2020},
	pages = {287--291},
}

@inproceedings{jhaver_bystanders_2024,
	address = {New York, NY, USA},
	series = {{CHI} '24},
	title = {Bystanders of {Online} {Moderation}: {Examining} the {Effects} of {Witnessing} {Post}-{Removal} {Explanations}},
	isbn = {979-8-4007-0330-0},
	url = {https://doi.org/10.1145/3613904.3642204},
	doi = {10.1145/3613904.3642204},
	booktitle = {Proceedings of the 2024 {CHI} {Conference} on {Human} {Factors} in {Computing} {Systems}},
	publisher = {Association for Computing Machinery},
	author = {Jhaver, Shagun and Rathi, Himanshu and Saha, Koustuv},
	year = {2024},
}

@article{jhaver_does_2019,
	address = {New York, NY, USA},
	title = {Does {Transparency} in {Moderation} {Really} {Matter}? {User} {Behavior} {After} {Content} {Removal} {Explanations} on {Reddit}},
	volume = {3},
	url = {https://doi.org/10.1145/3359252},
	doi = {10.1145/3359252},
	number = {CSCW},
	journal = {Proc. ACM Hum.-Comput. Interact.},
	publisher = {Association for Computing Machinery},
	author = {Jhaver, Shagun and Bruckman, Amy and Gilbert, Eric},
	month = nov,
	year = {2019},
}

@article{jhaver_evaluating_2021,
	address = {New York, NY, USA},
	title = {Evaluating the {Effectiveness} of {Deplatforming} as a {Moderation} {Strategy} on {Twitter}},
	volume = {5},
	url = {https://doi.org/10.1145/3479525},
	doi = {10.1145/3479525},
	number = {CSCW2},
	journal = {Proc. ACM Hum.-Comput. Interact.},
	publisher = {Association for Computing Machinery},
	author = {Jhaver, Shagun and Boylston, Christian and Yang, Diyi and Bruckman, Amy},
	month = oct,
	year = {2021},
}

@article{jhaver_did_2019,
	address = {New York, NY, USA},
	title = {"{Did} {You} {Suspect} the {Post} {Would} be {Removed}?": {Understanding} {User} {Reactions} to {Content} {Removals} on {Reddit}},
	volume = {3},
	url = {https://doi.org/10.1145/3359294},
	doi = {10.1145/3359294},
	number = {CSCW},
	journal = {Proc. ACM Hum.-Comput. Interact.},
	publisher = {Association for Computing Machinery},
	author = {Jhaver, Shagun and Appling, Darren Scott and Gilbert, Eric and Bruckman, Amy},
	month = nov,
	year = {2019},
}

@article{gerrard_beyond_2018,
	address = {London, England},
	title = {Beyond the hashtag: {Circumventing} content moderation on social media},
	volume = {20},
	issn = {1461-4448},
	doi = {10.1177/1461444818776611},
	number = {12},
	journal = {New media \& society},
	publisher = {SAGE Publications},
	author = {Gerrard, Ysabel},
	year = {2018},
	pages = {4492--4511},
}

@article{dillard_nature_2005,
	title = {On the {Nature} of {Reactance} and its {Role} in {Persuasive} {Health} {Communication}},
	volume = {72},
	issn = {0363-7751},
	url = {https://doi.org/10.1080/03637750500111815},
	doi = {10.1080/03637750500111815},
	number = {2},
	journal = {Communication Monographs},
	publisher = {Routledge},
	author = {Dillard, James Price and Shen, Lijiang},
	month = jun,
	year = {2005},
	pages = {144--168},
}

@article{deci_what_2000,
	title = {The "{What}" and "{Why}" of {Goal} {Pursuits}: {Human} {Needs} and the {Self}-{Determination} of {Behavior}},
	volume = {11},
	issn = {1047-840X},
	url = {https://doi.org/10.1207/S15327965PLI1104_01},
	doi = {10.1207/S15327965PLI1104_01},
	number = {4},
	journal = {Psychological Inquiry},
	publisher = {Routledge},
	author = {Deci, Edward L. and Ryan, Richard M.},
	month = oct,
	year = {2000},
	pages = {227--268},
}

@article{chandrasekharan_quarantined_2022,
	address = {New York, NY, USA},
	title = {Quarantined! {Examining} the {Effects} of a {Community}-{Wide} {Moderation} {Intervention} on {Reddit}},
	volume = {29},
	issn = {1073-0516},
	url = {https://doi.org/10.1145/3490499},
	doi = {10.1145/3490499},
	number = {4},
	journal = {ACM Trans. Comput.-Hum. Interact.},
	publisher = {Association for Computing Machinery},
	author = {Chandrasekharan, Eshwar and Jhaver, Shagun and Bruckman, Amy and Gilbert, Eric},
	month = mar,
	year = {2022},
}

@article{chandrasekharan_internets_2018,
	address = {New York, NY, USA},
	title = {The {Internet}'s {Hidden} {Rules}: {An} {Empirical} {Study} of {Reddit} {Norm} {Violations} at {Micro}, {Meso}, and {Macro} {Scales}},
	volume = {2},
	url = {https://doi.org/10.1145/3274301},
	doi = {10.1145/3274301},
	number = {CSCW},
	journal = {Proc. ACM Hum.-Comput. Interact.},
	publisher = {Association for Computing Machinery},
	author = {Chandrasekharan, Eshwar and Samory, Mattia and Jhaver, Shagun and Charvat, Hunter and Bruckman, Amy and Lampe, Cliff and Eisenstein, Jacob and Gilbert, Eric},
	month = nov,
	year = {2018},
}

@article{chandrasekharan_you_2017,
	address = {New York, NY, USA},
	title = {You {Can}'t {Stay} {Here}: {The} {Efficacy} of {Reddit}'s 2015 {Ban} {Examined} {Through} {Hate} {Speech}},
	volume = {1},
	url = {https://doi.org/10.1145/3134666},
	doi = {10.1145/3134666},
	number = {CSCW},
	journal = {Proc. ACM Hum.-Comput. Interact.},
	publisher = {Association for Computing Machinery},
	author = {Chandrasekharan, Eshwar and Pavalanathan, Umashanthi and Srinivasan, Anirudh and Glynn, Adam and Eisenstein, Jacob and Gilbert, Eric},
	month = dec,
	year = {2017},
}

@book{brown_politeness_1987,
	address = {Cambridge [Cambridgeshire] ;},
	series = {Studies in interactional sociolinguistics ; 4},
	title = {Politeness : some universals in language usage},
	isbn = {0-521-30862-3},
	publisher = {Cambridge University Press},
	author = {Brown, Penelope. and Levinson, Stephen C.},
	year = {1987},
}

@book{brehm_theory_1966,
	address = {New York},
	series = {Social psychology},
	title = {A theory of psychological reactance},
	publisher = {Academic Press},
	author = {Brehm, Jack Williams.},
	year = {1966},
}

@incollection{braithwaite_responsive_2001,
	title = {Responsive {Regulation}},
	isbn = {978-0-19-513639-5},
	url = {https://doi.org/10.1093/oso/9780195136395.003.0002},
	doi = {10.1093/oso/9780195136395.003.0002},
	urldate = {2026-05-30},
	booktitle = {Restorative {Justice} \& {Responsive} {Regulation}},
	publisher = {Oxford University Press},
	author = {Braithwaite, John},
	editor = {Braithwaite, John},
	month = nov,
	year = {2001},
	pages = {0},
}

@article{raiderbdev_reddit_nodate,
	title = {Reddit comments/submissions 2025-12},
	author = {{RaiderBDev}},
}

@inproceedings{binns_its_2018,
	address = {New York, NY, USA},
	series = {{CHI} '18},
	title = {'{It}'s {Reducing} a {Human} {Being} to a {Percentage}': {Perceptions} of {Justice} in {Algorithmic} {Decisions}},
	isbn = {978-1-4503-5620-6},
	url = {https://doi.org/10.1145/3173574.3173951},
	doi = {10.1145/3173574.3173951},
	booktitle = {Proceedings of the 2018 {CHI} {Conference} on {Human} {Factors} in {Computing} {Systems}},
	publisher = {Association for Computing Machinery},
	author = {Binns, Reuben and Van Kleek, Max and Veale, Michael and Lyngs, Ulrik and Zhao, Jun and Shadbolt, Nigel},
	year = {2018},
	pages = {1--14},
}

@incollection{sundar_toward_2015,
	title = {Toward a {Theory} of {Interactive} {Media} {Effects} ({TIME})},
	isbn = {978-1-118-42645-6},
	url = {https://onlinelibrary-wiley-com.libproxy2.usc.edu/doi/abs/10.1002/9781118426456.ch3},
	doi = {https://doi-org.libproxy2.usc.edu/10.1002/9781118426456.ch3},
	booktitle = {The {Handbook} of the {Psychology} of {Communication} {Technology}},
	publisher = {John Wiley \& Sons, Ltd},
	author = {Sundar, S. Shyam and Jia, Haiyan and Waddell, T. Franklin and Huang, Yan},
	year = {2015},
	note = {Section: 3
\_eprint: https://onlinelibrary-wiley-com.libproxy2.usc.edu/doi/pdf/10.1002/9781118426456.ch3},
	pages = {47--86},
}

@article{schauer_fear_1978,
	title = {Fear, risk and the first amendment: {Unraveling} the "chilling effect"},
	volume = {58},
	journal = {Boston University Law Review},
	author = {Schauer, Frederick},
	year = {1978},
	pages = {685--732},
}

@article{grimmelmann_virtues_2015,
	title = {The virtues of moderation},
	volume = {17},
	journal = {Yale Journal of Law \& Technology},
	author = {Grimmelmann, James},
	year = {2015},
	pages = {42--109},
}

@inproceedings{devito_algorithms_2017,
	address = {New York, NY, USA},
	series = {{CHI} '17},
	title = {"{Algorithms} ruin everything": \#{RIPTwitter}, {Folk} {Theories}, and {Resistance} to {Algorithmic} {Change} in {Social} {Media}},
	isbn = {978-1-4503-4655-9},
	url = {https://doi.org/10.1145/3025453.3025659},
	doi = {10.1145/3025453.3025659},
	booktitle = {Proceedings of the 2017 {CHI} {Conference} on {Human} {Factors} in {Computing} {Systems}},
	publisher = {Association for Computing Machinery},
	author = {DeVito, Michael Ann and Gergle, Darren and Birnholtz, Jeremy},
	year = {2017},
	pages = {3163--3174},
}

@article{christin_internal_2024,
	title = {Internal {Fractures}: {The} {Competing} {Logics} of {Social} {Media} {Platforms}},
	volume = {10},
	url = {https://doi.org/10.1177/20563051241274668},
	doi = {10.1177/20563051241274668},
	number = {3},
	journal = {Social Media + Society},
	author = {Christin, Angèle and Bernstein, Michael S. and Hancock, Jeffrey T. and Jia, Chenyan and Mado, Marijn N. and Tsai, Jeanne L. and Xu, Chunchen},
	year = {2024},
	note = {\_eprint: https://doi.org/10.1177/20563051241274668},
	pages = {20563051241274668},
}

@book{jenkins_convergence_2006,
	title = {Convergence {Culture}},
	isbn = {978-0-8147-4281-5},
	url = {http://www.jstor.org/stable/j.ctt9qffwr},
	urldate = {2026-04-10},
	publisher = {NYU Press},
	author = {Jenkins, Henry},
	year = {2006},
}

@article{baym_socially_2012,
	title = {Socially {Mediated} {Publicness}: {An} {Introduction}},
	volume = {56},
	url = {https://doi.org/10.1080/08838151.2012.705200},
	doi = {10.1080/08838151.2012.705200},
	number = {3},
	journal = {Journal of Broadcasting \& Electronic Media},
	publisher = {Routledge},
	author = {Baym, Nancy K. and boyd, danah},
	year = {2012},
	note = {\_eprint: https://doi.org/10.1080/08838151.2012.705200},
	pages = {320--329},
}

@article{walther_computer-mediated_1996,
	title = {Computer-{Mediated} {Communication}: {Impersonal}, {Interpersonal}, and {Hyperpersonal} {Interaction}},
	volume = {23},
	url = {https://doi.org/10.1177/009365096023001001},
	doi = {10.1177/009365096023001001},
	number = {1},
	journal = {Communication Research},
	author = {WALTHER, JOSEPH B.},
	year = {1996},
	note = {\_eprint: https://doi.org/10.1177/009365096023001001},
	pages = {3--43},
}

@article{sundar_rise_2020,
	title = {Rise of {Machine} {Agency}: {A} {Framework} for {Studying} the {Psychology} of {Human}–{AI} {Interaction} ({HAII})},
	volume = {25},
	issn = {1083-6101},
	url = {https://doi.org/10.1093/jcmc/zmz026},
	doi = {10.1093/jcmc/zmz026},
	number = {1},
	journal = {Journal of Computer-Mediated Communication},
	author = {Sundar, S Shyam},
	month = jan,
	year = {2020},
	note = {\_eprint: https://academic.oup.com/jcmc/article-pdf/25/1/74/32961171/zmz026.pdf},
	pages = {74--88},
}

@inproceedings{nass_computers_1994,
	address = {New York, NY, USA},
	series = {{CHI} '94},
	title = {Computers are social actors},
	isbn = {0-89791-650-6},
	url = {https://doi.org/10.1145/191666.191703},
	doi = {10.1145/191666.191703},
	booktitle = {Proceedings of the {SIGCHI} {Conference} on {Human} {Factors} in {Computing} {Systems}},
	publisher = {Association for Computing Machinery},
	author = {Nass, Clifford and Steuer, Jonathan and Tauber, Ellen R.},
	year = {1994},
	pages = {72--78},
}

@article{srinivasan_content_2019,
	address = {New York, NY, USA},
	title = {Content {Removal} as a {Moderation} {Strategy}: {Compliance} and {Other} {Outcomes} in the {ChangeMyView} {Community}},
	volume = {3},
	url = {https://doi.org/10.1145/3359265},
	doi = {10.1145/3359265},
	number = {CSCW},
	journal = {Proc. ACM Hum.-Comput. Interact.},
	publisher = {Association for Computing Machinery},
	author = {Srinivasan, Kumar Bhargav and Danescu-Niculescu-Mizil, Cristian and Lee, Lillian and Tan, Chenhao},
	month = nov,
	year = {2019},
}

@article{horta_ribeiro_platform_2021,
	address = {New York, NY, USA},
	title = {Do {Platform} {Migrations} {Compromise} {Content} {Moderation}? {Evidence} from r/{The}\_Donald and r/{Incels}},
	volume = {5},
	url = {https://doi.org/10.1145/3476057},
	doi = {10.1145/3476057},
	number = {CSCW2},
	journal = {Proc. ACM Hum.-Comput. Interact.},
	publisher = {Association for Computing Machinery},
	author = {Horta Ribeiro, Manoel and Jhaver, Shagun and Zannettou, Savvas and Blackburn, Jeremy and Stringhini, Gianluca and De Cristofaro, Emiliano and West, Robert},
	month = oct,
	year = {2021},
}

@book{gillespie_custodians_2019,
	title = {Custodians of the {Internet}: {Platforms}, {Content} {Moderation}, and the {Hidden} {Decisions} {That} {Shape} {Social} {Media}},
	isbn = {978-0-300-23502-9},
	shorttitle = {Custodians of the {Internet}},
	url = {https://www.degruyter.com/document/doi/10.12987/9780300235029/html},
	doi = {10.12987/9780300235029},
	language = {en},
	urldate = {2024-09-20},
	publisher = {Yale University Press},
	author = {Gillespie, Tarleton},
	month = dec,
	year = {2019},
}

@article{myers_west_censored_2018,
	title = {Censored, suspended, shadowbanned: {User} interpretations of content moderation on social media platforms},
	volume = {20},
	issn = {1461-4448, 1461-7315},
	shorttitle = {Censored, suspended, shadowbanned},
	url = {http://journals.sagepub.com/doi/10.1177/1461444818773059},
	doi = {10.1177/1461444818773059},
	language = {en},
	number = {11},
	urldate = {2024-09-20},
	journal = {New Media \& Society},
	author = {Myers West, Sarah},
	month = nov,
	year = {2018},
	pages = {4366--4383},
}

@article{goncalves_common_2023,
	title = {Common sense or censorship: {How} algorithmic moderators and message type influence perceptions of online content deletion},
	volume = {25},
	url = {https://doi.org/10.1177/14614448211032310},
	doi = {10.1177/14614448211032310},
	number = {10},
	journal = {New Media \& Society},
	author = {Gonçalves, João and Weber, Ina and Masullo, Gina M. and Silva, Marisa Torres da and Hofhuis, Joep},
	year = {2023},
	note = {\_eprint: https://doi.org/10.1177/14614448211032310},
	pages = {2595--2617},
}

@article{gillespie_not_2022,
	title = {Do {Not} {Recommend}? {Reduction} as a {Form} of {Content} {Moderation}},
	volume = {8},
	url = {https://doi.org/10.1177/20563051221117552},
	doi = {10.1177/20563051221117552},
	number = {3},
	journal = {Social Media + Society},
	author = {Gillespie, Tarleton},
	year = {2022},
	note = {\_eprint: https://doi.org/10.1177/20563051221117552},
	pages = {20563051221117552},
}

@techreport{caplan_content_2018,
	address = {New York},
	title = {Content or {Context} {Moderation}? {Artisanal}, {Community}, and {Industrial} {Approaches}},
	url = {https://datasociety.net/library/content-or-context-moderation/},
	institution = {Data \& Society Research Institute},
	author = {Caplan, Robyn},
	year = {2018},
}

@article{saleem_aftermath_2018,
	title = {The aftermath of disbanding an online hateful community},
	journal = {arXiv preprint arXiv:1804.07354},
	author = {Saleem, Haji Mohammad and Ruths, Derek},
	year = {2018},
}

@article{molina_when_2022,
	title = {When {AI} moderates online content: effects of human collaboration and interactive transparency on user trust},
	volume = {27},
	copyright = {https://creativecommons.org/licenses/by/4.0/},
	issn = {1083-6101},
	shorttitle = {When {AI} moderates online content},
	url = {https://academic.oup.com/jcmc/article/doi/10.1093/jcmc/zmac010/6648459},
	doi = {10.1093/jcmc/zmac010},
	language = {en},
	number = {4},
	urldate = {2024-09-20},
	journal = {Journal of Computer-Mediated Communication},
	author = {Molina, Maria D and Sundar, S Shyam},
	editor = {Lee, Eun-Ju},
	month = jul,
	year = {2022},
	pages = {zmac010},
}

@article{ferrara_rise_2016,
	title = {The rise of social bots},
	volume = {59},
	issn = {0001-0782, 1557-7317},
	url = {https://dl.acm.org/doi/10.1145/2818717},
	doi = {10.1145/2818717},
	language = {en},
	number = {7},
	urldate = {2024-09-20},
	journal = {Communications of the ACM},
	author = {Ferrara, Emilio and Varol, Onur and Davis, Clayton and Menczer, Filippo and Flammini, Alessandro},
	month = jun,
	year = {2016},
	pages = {96--104},
}
\newpage
\onecolumn
\section{Appendix}
\section{Data Overview}
\begin{table*}[htbp]
\centering
\caption{ Summary Statistics of Moderator Roles and Activity}
\begin{tabular}{lrrr}
\toprule
\textbf{Metric} & \textbf{Bot} & \textbf{Modteam} & \textbf{Personal Accounts} \\
\midrule
Unique Moderator Accounts & 967 & 50,689 & 80,121 \\
Total Explanations & 10,163,786 & 557,065 & 1,074,185 \\
Unique Users Moderated & 7,748,301 & 534,446 & 1,002,663 \\
Unique Subreddits & 51,738 & 1,559 & 7,964 \\
\bottomrule
\end{tabular}
\label{tab:moderation_summary_transposed}
\end{table*}

\begin{table*}[htbp]
\centering
\caption{Summary statistics by moderation reason and moderator type (alphabetical order)}
\label{tab:moderation_reason_summary_alpha}
\resizebox{\textwidth}{!}{%
\begin{tabular}{lrrrrrrrrr}
\hline
& \multicolumn{3}{c}{Bot} & \multicolumn{3}{c}{Modteam} & \multicolumn{3}{c}{Personal} \\
\cline{2-4} \cline{5-7} \cline{8-10}
Reason & Unique Users & Unique Subreddits & Counts & Unique Users & Unique Subreddits & Counts & Unique Users & Unique Subreddits & Counts \\
\hline
Account Issues              & 4,554,722 & 2,715 & 5,601,262 & 102,322 & 366 & 102,854 & 162,954 & 1,629 & 165,473 \\
Content Quality Issues      & 499,241   & 878  & 529,658   & 91,384  & 356 & 91,991  & 169,482 & 1,166 & 171,536 \\
Formatting or Structure     & 1,695,014 & 1,233 & 1,814,501 & 88,647  & 368 & 90,209  & 144,526 & 1,252 & 147,187 \\
Harmful or Illegal Content  & 47,699    & 302  & 51,503    & 38,275  & 453 & 38,982  & 40,172  & 1,032 & 40,628 \\
Memeber Reports             & 10,073    & 2    & 10,073    & 11,082  & 10  & 11,106  & 2       & 2     & 2 \\
Rule or Scope Violations    & 753,706   & 850  & 783,437   & 160,040 & 490 & 162,511 & 268,174 & 1,866 & 274,325 \\
Spam or Scams               & 263,004   & 784  & 305,503   & 43,645  & 430 & 43,997  & 184,262 & 1,524 & 188,480 \\
\hline
\end{tabular}%
}
\end{table*}

\begin{table*}[htbp]
\centering
\caption{Distribution of Removal Reasons (keyword based) Across Subreddits}
\label{tab:removal_reasons_distribution}
\resizebox{\textwidth}{!}{
\begin{tabular}{lrr}
\hline
Removal reason & Post count & Subreddit count \\
\hline
Not enough karma & 87,123 & 310 \\
Not verified & 37,972 & 69 \\
New account & 24,266 & 327 \\
Wrong format & 21,732 & 212 \\
Spam & 13,394 & 121 \\
Exclude & 9,149 & 52 \\
Begging or pandering & 8,908 & 48 \\
Prohibited content & 8,129 & 159 \\
Lack of user flair & 6,422 & 39 \\
Other & 5,385 & 330 \\
Insufficiently elaborated / low-quality post & 5,162 & 82 \\
Questions not allowed & 3,371 & 49 \\
Lack of source or potential misinformation & 2,159 & 70 \\
Exceeds limit & 1,350 & 47 \\
Non-standard character in post & 1,238 & 123 \\
Clickbait & 958 & 53 \\
Off-topic & 818 & 4 \\
Title irrelevant to content & 601 & 18 \\
Upvote baiting & 528 & 42 \\
Username not allowed in post & 521 & 35 \\
Post body not allowed & 485 & 2 \\
Outsourcing not allowed & 394 & 10 \\
Admission questions not allowed & 330 & 2 \\
Missing required content & 258 & 8 \\
Content only allowed in comment & 242 & 2 \\
Member reports & 229 & 3 \\
Not following specific subreddit rules & 200 & 1 \\
Political content not allowed & 194 & 1 \\
Harmful or malicious content & 189 & 33 \\
Violation not specified & 170 & 3 \\
Copyright infringement & 130 & 4 \\
Repetitive or generic content & 121 & 4 \\
Sex with minor & 42 & 3 \\
Fake contest & 16 & 2 \\
Sexting & 6 & 4 \\
\hline
\end{tabular}
}
\end{table*}

\begin{table*}[htbp]
\centering
\caption{Grouping of Moderation Reasons into Analytical Categories}
\label{tab:reason_grouping}
\resizebox{\textwidth}{!}{
\begin{tabular}{lp{12cm}}
\hline
\textbf{Analytical category} & \textbf{Underlying moderation reasons} \\
\hline
Account issues &
Not enough karma; Not verified; New account \\

Formatting or structure &
Wrong format; Exceeds limit; Non-standard character in post; 
Title irrelevant to content; Username not allowed in post; 
Post body not allowed; Missing required content; 
Content only allowed in comment; Lack of user flair \\

Spam or scams &
Spam; Begging or pandering; Clickbait; Upvote baiting; 
Outsourcing not allowed; Fake contest; Repetitive or generic content \\

Rule or scope violations &
Questions not allowed; Admission questions not allowed; 
Political content not allowed; Prohibited content; 
Off-topic; Not following specific subreddit rules; 
Violation not specified \\

Content quality issues &
Insufficiently elaborated post or low-quality post, Lack of source or potential misinformation \\

Harmful or illegal content &
Harmful or malicious content; Copyright infringement; 
Sex with minor; Sexting \\
\hline
\end{tabular}
}
\end{table*}
\section{Method}
\subsection{Distribution of Baseline user activity}

\begin{figure}[h]
    \centering
    \begin{subfigure}[t]{0.32\textwidth}
        \includegraphics[width=\textwidth]{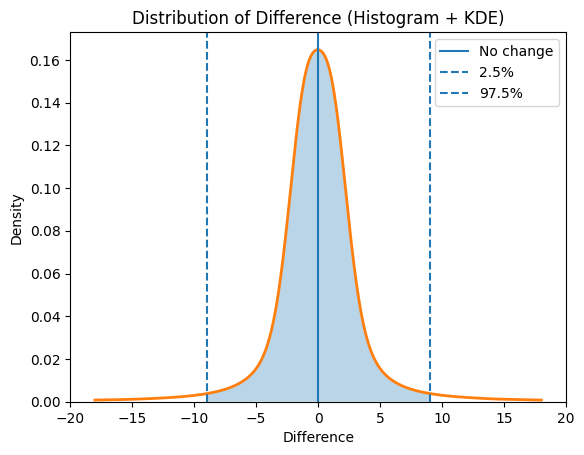}
        \caption{Difference}
        \label{fig:post_freq}
    \end{subfigure}
    \hfill
    \begin{subfigure}[t]{0.32\textwidth}
        \includegraphics[width=\textwidth]{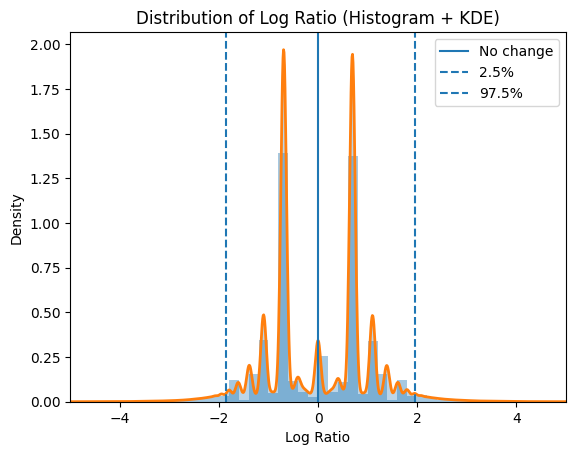}
        \caption{Log Ratio}
        \label{fig:resistant}
    \end{subfigure}
    \hfill
    \begin{subfigure}[t]{0.32\textwidth}
        \includegraphics[width=\textwidth]{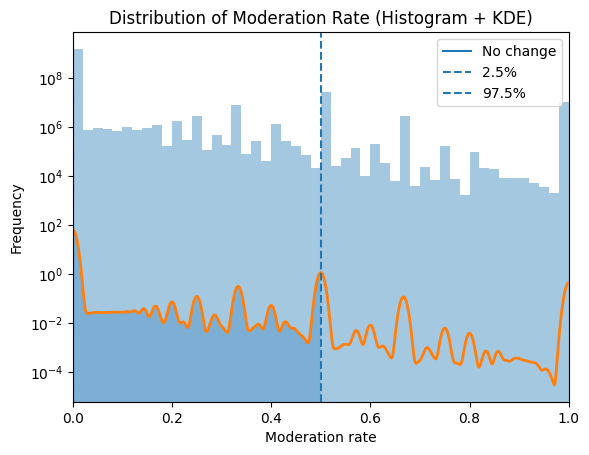}
        \caption{Moderation Rate}
        \label{fig:compliance}
    \end{subfigure}
    \caption{Distribution for difference of post frequency, log ratio of post frequency, and moderation rate. log ratio: $n = 2670323207, \mu = 0.0015 , \sigma = 1.023 , skew = 0.062, kurtosis = 0.29$, moderation rate: $n = 2670323207, \mu = 0.020 , \sigma = ? , skew = 6.42, kurtosis = 44.8$ }
    \label{fig:distribution}
\end{figure}

\newpage
\subsection{Principle Component Selection}
\small
\setlength{\tabcolsep}{5pt}
\begin{longtable}{@{}lp{4.5cm}cp{4cm}p{4cm}@{}}
\caption{Principal components, interpretive themes, and top feature loadings.
  Each PC is labelled with its theme and variance explained. The five
  features with the largest absolute loadings are listed by direction.
  Feature names follow LIWC-22 and custom dictionary conventions.
  Positive/negative signs indicate the direction of the association.}
\label{tab:pc_summary} \\
\toprule
\textbf{PC} & \textbf{Theme} & \textbf{Var.\,(\%)} & \textbf{Positive loadings} & \textbf{Negative loadings} \\
\midrule
\multicolumn{5}{@{}l}{\textit{\textbf{Child-level PCs (CH)}}} \\[2pt]
\texttt{CH\_PC1} & Spatial Detachment vs Polite Reflective Engagement & 10.7 & \textit{space} (+0.31), \textit{we} (+0.26) & \textit{focuspast} (-0.29), \textit{tentat} (-0.29), \textit{polite} (-0.28) \\
\texttt{CH\_PC2} & Direct Positive Address & 8.3 & \textit{you} (+0.36), \textit{tone\_pos} (+0.34), \textit{comm} (+0.33), \textit{socrefs} (+0.33), \textit{we} (+0.29) & --- \\
\texttt{CH\_PC3} & Negative Tentative Tone with Comma-Pacing & 6.5 & \textit{Comma} (+0.33), \textit{tentat} (+0.30), \textit{tone\_neg} (+0.30), \textit{differ} (+0.29) & \textit{tech} (-0.31) \\
\texttt{CH\_PC4} & Work-Time Framing vs Causal Power & 6.4 & \textit{work} (+0.29), \textit{differ} (+0.28), \textit{time} (+0.27) & \textit{power} (-0.27), \textit{cause} (-0.26) \\
\texttt{CH\_PC5} & Explanatory Causal Reasoning & 5.3 & \textit{cause} (+0.34), \textit{motion} (+0.33), \textit{insight} (+0.29), \textit{allure} (+0.25) & \textit{socrefs} (-0.29) \\
\texttt{CH\_PC6} & Contracted Need Expression vs Visual Attentional & 5.0 & \textit{Apostro} (+0.36), \textit{need} (+0.35) & \textit{attention} (-0.35), \textit{visual} (-0.31), \textit{focuspresent} (-0.29) \\
\texttt{CH\_PC7} & Achievement Positive Framing & 4.4 & \textit{achieve} (+0.62), \textit{work} (+0.35), \textit{tone\_pos} (+0.28), \textit{OtherP} (+0.24) & \textit{need} (-0.22) \\
\texttt{CH\_PC8} & Reflective Uncertainty & 3.9 & \textit{discrep} (+0.38), \textit{netspeak} (+0.33), \textit{insight} (+0.32) & \textit{time} (-0.32), \textit{power} (-0.28) \\
\texttt{CH\_PC9} & Emphatic Prosocial Exclamation & 3.7 & \textit{Exclam} (+0.45), \textit{prosocial} (+0.28) & \textit{cause} (-0.32), \textit{power} (-0.32), \textit{you} (-0.25) \\
\texttt{CH\_PC10} & Future vs Present Temporal Focus & 3.4 & \textit{focusfuture} (+0.45), \textit{attention} (+0.34), \textit{you} (+0.25) & \textit{focuspresent} (-0.39), \textit{work} (-0.29) \\
\texttt{CH\_PC11} & Absolute All-or-Nothing Language & 3.0 & \textit{allnone} (+0.49), \textit{focusfuture} (+0.30), \textit{tentat} (+0.22) & \textit{attention} (-0.35), \textit{Exclam} (-0.22) \\
\texttt{CH\_PC12} & Comma-Paced Allure vs Discrepancy Need & 3.0 & \textit{Comma} (+0.51), \textit{allure} (+0.32) & \textit{discrep} (-0.37), \textit{allnone} (-0.26), \textit{need} (-0.23) \\
\texttt{CH\_PC13} & Punctuated Visual-Temporal Description & 2.9 & \textit{Period} (+0.40), \textit{visual} (+0.33), \textit{OtherP} (+0.31), \textit{time} (+0.30), \textit{allure} (+0.28) & --- \\
\texttt{CH\_PC14} & Emphatic Negative Past-Focused Reflection & 2.6 & \textit{Exclam} (+0.39), \textit{tone\_neg} (+0.38), \textit{focuspast} (+0.31), \textit{netspeak} (+0.29) & \textit{discrep} (-0.28) \\
\texttt{CH\_PC15} & Emphatic Causal Exclamation & 2.6 & \textit{Exclam} (+0.48), \textit{cause} (+0.34) & \textit{attention} (-0.38), \textit{tech} (-0.29), \textit{need} (-0.24) \\
\texttt{CH\_PC16} & Temporal Framing vs Punctuated Present & 2.5 & \textit{time} (+0.51), \textit{netspeak} (+0.28) & \textit{Comma} (-0.36), \textit{Period} (-0.34), \textit{focuspresent} (-0.31) \\
\texttt{CH\_PC17} & Visual Netspeak vs Formal OtherP-Time & 2.3 & \textit{visual} (+0.39), \textit{netspeak} (+0.37), \textit{differ} (+0.28) & \textit{OtherP} (-0.28), \textit{time} (-0.28) \\
\texttt{CH\_PC18} & Present-Focus vs Causal-Need Future & 2.2 & \textit{focuspresent} (+0.46), \textit{focusfuture} (+0.28) & \textit{visual} (-0.37), \textit{need} (-0.28), \textit{cause} (-0.27) \\
\texttt{CH\_PC19} & Differentiation vs Prosocial Negative Tone & 2.1 & \textit{differ} (+0.48), \textit{comm} (+0.32), \textit{OtherP} (+0.27) & \textit{tone\_neg} (-0.34), \textit{prosocial} (-0.33) \\
[4pt]
\multicolumn{5}{@{}l}{\textit{\textbf{Header-level PCs (HL)}}} \\[2pt]
\texttt{HL\_PC1} & Conversational vs Analytical Register & 19.6 & \textit{function} (+0.36), \textit{pronoun} (+0.32), \textit{Dic} (+0.32), \textit{Cognition} (+0.30) & \textit{Analytic} (-0.32) \\
\texttt{HL\_PC2} & Motivational Content Richness & 11.0 & \textit{prep} (+0.36), \textit{Drives} (+0.33), \textit{Affect} (+0.31), \textit{Perception} (+0.30), \textit{WC} (+0.26) & --- \\
\texttt{HL\_PC3} & Warm Social Tone & 8.9 & \textit{Social} (+0.34), \textit{Clout} (+0.32), \textit{Tone} (+0.30), \textit{Affect} (+0.28) & \textit{article} (-0.27) \\
\texttt{HL\_PC4} & Concrete Noun-Heavy Specificity & 7.4 & \textit{article} (+0.43), \textit{det} (+0.41), \textit{Clout} (+0.32) & \textit{adverb} (-0.28), \textit{Authentic} (-0.26) \\
\texttt{HL\_PC5} & Quantified Direct Content & 6.9 & \textit{quantity} (+0.40), \textit{Lifestyle} (+0.33), \textit{adj} (+0.31) & \textit{adverb} (-0.37), \textit{prep} (-0.29) \\
\texttt{HL\_PC6} & Rule-Prohibitive Negation & 5.1 & \textit{negate} (+0.37), \textit{Drives} (+0.27) & \textit{Culture} (-0.35), \textit{Authentic} (-0.29), \textit{Lifestyle} (-0.29) \\
\texttt{HL\_PC7} & Numerical Authority & 4.4 & \textit{number} (+0.43), \textit{Clout} (+0.32), \textit{quantity} (+0.32), \textit{pronoun} (+0.29) & \textit{article} (-0.31) \\
\texttt{HL\_PC8} & Message Length and Elaboration & 3.7 & \textit{WPS} (+0.44), \textit{WC} (+0.41), \textit{Conversation} (+0.28), \textit{negate} (+0.28), \textit{Authentic} (+0.27) & --- \\
\texttt{HL\_PC9} & Community-Scale Numerical Reference & 3.4 & \textit{number} (+0.46), \textit{Social} (+0.37), \textit{Dic} (+0.23) & \textit{Lifestyle} (-0.38), \textit{ipron} (-0.35) \\
\texttt{HL\_PC10} & Netspeak Informality vs Cultural Formality & 3.0 & \textit{Conversation} (+0.45), \textit{adj} (+0.25) & \textit{Culture} (-0.41), \textit{AllPunc} (-0.29), \textit{ipron} (-0.25) \\
\texttt{HL\_PC11} & Logical Connective Elaboration & 2.7 & \textit{conj} (+0.45), \textit{Drives} (+0.28) & \textit{verb} (-0.31), \textit{Conversation} (-0.28), \textit{Culture} (-0.28) \\
\texttt{HL\_PC12} & Descriptive Positive Framing & 2.5 & \textit{adj} (+0.53), \textit{Authentic} (+0.25), \textit{Drives} (+0.24) & \textit{negate} (-0.34), \textit{Lifestyle} (-0.31) \\
\texttt{HL\_PC13} & Adverbial Hedging vs Structured Perception & 2.5 & \textit{adverb} (+0.54) & \textit{Perception} (-0.36), \textit{conj} (-0.33), \textit{verb} (-0.28), \textit{Conversation} (-0.27) \\
\bottomrule

\end{longtable}

\newpage
\section{Result}
\subsection{ANOVA and Dunn post-hoc analysis on different behavioral trjactories when moderated by different source }
\section*{Appendix: Statistical Tests by Metric}

% ============================================================
\subsection*{Metric: Self-Censor}
% ============================================================

The Kruskal-Wallis test revealed a significant difference across moderator groups
($H = 663{,}237.06$, $p < .0001$). Post-hoc Dunn tests with Bonferroni correction
confirmed that all pairwise comparisons were significant ($p < .0001$).

\begin{table*}[H]
\centering
\caption{Tukey HSD Pairwise Comparisons -- Self-Censor}
\begin{tabular}{llrrrrl}
\toprule
\textbf{Group 1} & \textbf{Group 2} & \textbf{Mean Diff} & \textbf{$p$-adj} & \textbf{Lower} & \textbf{Upper} & \textbf{Reject $H_0$} \\
\midrule
bots & modteam          &  0.0849 & $<.0001$ & 0.0846 &  0.0852 & Yes \\
bots & personalAccounts &  0.0724 & $<.0001$ & 0.0722 &  0.0726 & Yes \\
modteam & personalAccounts & $-$0.0125 & $<.0001$ & $-$0.0128 & $-$0.0121 & Yes \\
\bottomrule
\end{tabular}
\end{table*}

\begin{table*}[H]
\centering
\caption{Self-censorship Pairwise Differences}

\vspace{0.5em}

\begin{tabular}{llrl}
\toprule
\textbf{Comparison} & \textbf{Median Diff} & \textbf{$p$-value} & \textbf{Sig.} \\
\midrule
bots vs.\ modteam          & $-$0.0942 & $<.0001$ & * \\
bots vs.\ personalAccounts & $-$0.0556 & $<.0001$ & * \\
modteam vs.\ personalAccounts & $+$0.0386 & $<.0001$ & * \\
\bottomrule
\end{tabular}
\end{table*}

% ============================================================
\subsection*{Metric: Resistance}
% ============================================================

The Kruskal-Wallis test revealed a significant difference across moderator groups
($H = 146{,}970.87$, $p < .0001$). Post-hoc Dunn tests with Bonferroni correction
confirmed that all pairwise comparisons were significant ($p < .0001$).

\begin{table*}[H]
\centering
\caption{Tukey HSD Pairwise Comparisons -- Resistance}
\begin{tabular}{llrrrrl}
\toprule
\textbf{Group 1} & \textbf{Group 2} & \textbf{Mean Diff} & \textbf{$p$-adj} & \textbf{Lower} & \textbf{Upper} & \textbf{Reject $H_0$} \\
\midrule
bots & modteam          & $-$0.0042 & $<.0001$ & $-$0.0047 & $-$0.0036 & Yes \\
bots & personalAccounts & $-$0.0263 & $<.0001$ & $-$0.0267 & $-$0.0259 & Yes \\
modteam & personalAccounts & $-$0.0222 & $<.0001$ & $-$0.0228 & $-$0.0215 & Yes \\
\bottomrule
\end{tabular}
\end{table*}

\begin{table*}[H]
\centering
\caption{Resistance Pairwise Differences }

\vspace{0.5em}

\begin{tabular}{llrl}
\toprule
\textbf{Comparison} & \textbf{Median Diff} & \textbf{$p$-value} & \textbf{Sig.} \\
\midrule
bots vs.\ modteam          & $+$0.0000 & $<.0001$ & * \\
bots vs.\ personalAccounts & $+$0.0000 & $<.0001$ & * \\
modteam vs.\ personalAccounts & $+$0.0000 & $5.42 \times 10^{-238}$ & * \\
\bottomrule
\end{tabular}
\end{table*}

\noindent\textit{Note: All groups share a median of 0; significant differences detected by the Kruskal-Wallis and Dunn tests reflect differences in distributional shape and higher moments rather than central tendency.}

% ============================================================
\subsection*{Metric: Compliance}
% ============================================================

The Kruskal-Wallis test revealed a significant difference across moderator groups
($H = 330{,}201.54$, $p < .0001$). Post-hoc Dunn tests with Bonferroni correction
confirmed that all pairwise comparisons were significant ($p < .0001$).

\begin{table*}[H]
\centering
\caption{Tukey HSD Pairwise Comparisons -- Compliance}
\begin{tabular}{llrrrrl}
\toprule
\textbf{Group 1} & \textbf{Group 2} & \textbf{Mean Diff} & \textbf{$p$-adj} & \textbf{Lower} & \textbf{Upper} & \textbf{Reject $H_0$} \\
\midrule
bots & modteam          & $-$0.0778 & $<.0001$ & $-$0.0784 & $-$0.0773 & Yes \\
bots & personalAccounts & $-$0.0445 & $<.0001$ & $-$0.0449 & $-$0.0441 & Yes \\
modteam & personalAccounts &  0.0333 & $<.0001$ &  0.0327 &  0.0340 & Yes \\
\bottomrule
\end{tabular}
\end{table*}

\begin{table*}[H]
\centering
\caption{Compliance Pairwise Differences}

\vspace{0.5em}

\begin{tabular}{llrl}
\toprule
\textbf{Comparison} & \textbf{Median Diff} & \textbf{$p$-value} & \textbf{Sig.} \\
\midrule
bots vs.\ modteam          & $+$0.0941 & $<.0001$ & * \\
bots vs.\ personalAccounts & $+$0.0941 & $<.0001$ & * \\
modteam vs.\ personalAccounts & $+$0.0000 & $<.0001$ & * \\
\bottomrule
\end{tabular}
\end{table*}

\section{Linguistic regression results}

\section*{A\quad Main Effects Results}

The following tables report logistic regression main effects for each
outcome (Self-Censor, Resistance, Compliance) and each PCA set (HL: High-Level,
CH: Children). Columns show the coefficient ($\hat\beta$), standard error,
$z$-statistic, $p$-value, 95\% confidence interval, and FDR-adjusted significance.
$\checkmark$ denotes FDR $q<.05$ for linguistic PCs (reported where applicable).

\begin{table*}[H]
\centering
\small
\caption{Main effects: High-Level (HL), Self-Censor. Linguistic PCs use FDR-adjusted $q$; moderator terms use nominal $p$.}
\label{tab:main_HL_selfcensor_prob}
\begin{tabular}{llrrrrrrc}
\toprule
\textbf{Type} & \textbf{Term} & $\hat{\beta}$ & \textbf{SE} & $z$ & $p$ & \textbf{95\% CI low} & \textbf{95\% CI high} & \textbf{Sig.} \\
\midrule
Intercept & Intercept & -2.3204 & 0.0254 & -91.513 & $<.001$ & -2.3701 & -2.2707 &  \\
\midrule
Moderator Type & modteam & 0.5312 & 0.0662 & 8.028 & $<.001$ & 0.4015 & 0.6609 & \checkmark \\
 & personalAccounts & 0.4321 & 0.0649 & 6.659 & $<.001$ & 0.3049 & 0.5593 & \checkmark \\
\midrule
Moderation Reason & content\_quality\_issues & 0.0736 & 0.0463 & 1.590 & $0.112$ & -0.0171 & 0.1643 &  \\
 & formatting\_or\_structure & -0.1821 & 0.0461 & -3.948 & $<.001$ & -0.2725 & -0.0917 & \checkmark \\
 & harmful\_or\_illegal\_content & 0.1401 & 0.1029 & 1.361 & $0.173$ & -0.0616 & 0.3418 &  \\
 & rule\_or\_scope\_violations & 0.0276 & 0.0616 & 0.448 & $0.654$ & -0.0931 & 0.1483 &  \\
\bottomrule
\end{tabular}
\end{table*}

\begin{table*}[H]
\centering
\small
\caption{Main effects: High-Level (HL), Resistance. Linguistic PCs use FDR-adjusted $q$; moderator terms use nominal $p$.}
\label{tab:main_HL_resistance_prob}
\begin{tabular}{llrrrrrrc}
\toprule
\textbf{Type} & \textbf{Term} & $\hat{\beta}$ & \textbf{SE} & $z$ & $p$ & \textbf{95\% CI low} & \textbf{95\% CI high} & \textbf{Sig.} \\
\midrule
Intercept & Intercept & -3.1952 & 0.1257 & -25.428 & $<.001$ & -3.4415 & -2.9489 &  \\
\midrule
Moderator Type & modteam & 0.5968 & 0.2583 & 2.310 & $0.021$ & 0.0905 & 1.1032 & \checkmark \\
 & personalAccounts & 0.0663 & 0.2644 & 0.251 & $0.802$ & -0.4520 & 0.5846 &  \\
\midrule
Moderation Reason & content\_quality\_issues & -0.7201 & 0.2305 & -3.124 & $0.002$ & -1.1719 & -0.2683 & \checkmark \\
 & formatting\_or\_structure & -0.4549 & 0.1829 & -2.487 & $0.013$ & -0.8134 & -0.0964 & \checkmark \\
 & harmful\_or\_illegal\_content & 0.5479 & 0.5196 & 1.054 & $0.292$ & -0.4705 & 1.5662 &  \\
 & rule\_or\_scope\_violations & -0.2725 & 0.2198 & -1.240 & $0.215$ & -0.7033 & 0.1583 &  \\
\bottomrule
\end{tabular}
\end{table*}

\begin{table*}[H]
\centering
\small
\caption{Main effects: High-Level (HL), Compliance. Linguistic PCs use FDR-adjusted $q$; moderator terms use nominal $p$.}
\label{tab:main_HL_compliance_prob}
\begin{tabular}{llrrrrrrc}
\toprule
\textbf{Type} & \textbf{Term} & $\hat{\beta}$ & \textbf{SE} & $z$ & $p$ & \textbf{95\% CI low} & \textbf{95\% CI high} & \textbf{Sig.} \\
\midrule
Intercept & Intercept & 1.7303 & 0.0347 & 49.914 & $<.001$ & 1.6624 & 1.7983 &  \\
\midrule
Moderator Type & modteam & -0.6944 & 0.0736 & -9.436 & $<.001$ & -0.8387 & -0.5502 & \checkmark \\
 & personalAccounts & -0.4516 & 0.0681 & -6.635 & $<.001$ & -0.5850 & -0.3182 & \checkmark \\
\midrule
Moderation Reason & content\_quality\_issues & 0.2230 & 0.0581 & 3.837 & $<.001$ & 0.1091 & 0.3369 & \checkmark \\
 & formatting\_or\_structure & 0.3459 & 0.0615 & 5.622 & $<.001$ & 0.2253 & 0.4665 & \checkmark \\
 & harmful\_or\_illegal\_content & -0.2531 & 0.2197 & -1.152 & $0.249$ & -0.6837 & 0.1775 &  \\
 & rule\_or\_scope\_violations & 0.1015 & 0.0752 & 1.350 & $0.177$ & -0.0459 & 0.2488 &  \\
\bottomrule
\end{tabular}
\end{table*}

\begin{table*}[H]
\centering
\small
\caption{Main effects: Children (CH), Self-Censor. Linguistic PCs use FDR-adjusted $q$; moderator terms use nominal $p$.}
\label{tab:main_CH_selfcensor_prob}
\begin{tabular}{llrrrrrrc}
\toprule
\textbf{Type} & \textbf{Term} & $\hat{\beta}$ & \textbf{SE} & $z$ & $p$ & \textbf{95\% CI low} & \textbf{95\% CI high} & \textbf{Sig.} \\
\midrule
Intercept & Intercept & -2.2845 & 0.0237 & -96.531 & $<.001$ & -2.3309 & -2.2381 &  \\
\midrule
Moderator Type & modteam & 0.5800 & 0.0662 & 8.755 & $<.001$ & 0.4501 & 0.7098 & \checkmark \\
 & personalAccounts & 0.4478 & 0.0600 & 7.467 & $<.001$ & 0.3303 & 0.5654 & \checkmark \\
\midrule
Moderation Reason & content\_quality\_issues & -0.0026 & 0.0473 & -0.056 & $0.956$ & -0.0954 & 0.0901 &  \\
 & formatting\_or\_structure & -0.2592 & 0.0544 & -4.769 & $<.001$ & -0.3657 & -0.1527 & \checkmark \\
 & harmful\_or\_illegal\_content & -0.0782 & 0.1051 & -0.744 & $0.457$ & -0.2842 & 0.1278 &  \\
 & rule\_or\_scope\_violations & 0.0088 & 0.0539 & 0.163 & $0.870$ & -0.0969 & 0.1145 &  \\
\bottomrule
\end{tabular}
\end{table*}

\begin{table*}[H]
\centering
\small
\caption{Main effects: Children (CH), Resistance. Linguistic PCs use FDR-adjusted $q$; moderator terms use nominal $p$.}
\label{tab:main_CH_resistance_prob}
\begin{tabular}{llrrrrrrc}
\toprule
\textbf{Type} & \textbf{Term} & $\hat{\beta}$ & \textbf{SE} & $z$ & $p$ & \textbf{95\% CI low} & \textbf{95\% CI high} & \textbf{Sig.} \\
\midrule
Intercept & Intercept & -3.0098 & 0.1280 & -23.510 & $<.001$ & -3.2608 & -2.7589 &  \\
\midrule
Moderator Type & modteam & 0.2926 & 0.3458 & 0.846 & $0.397$ & -0.3851 & 0.9703 &  \\
 & personalAccounts & -0.0939 & 0.3722 & -0.252 & $0.801$ & -0.8234 & 0.6357 &  \\
\midrule
Moderation Reason & content\_quality\_issues & -0.6535 & 0.2090 & -3.126 & $0.002$ & -1.0632 & -0.2438 & \checkmark \\
 & formatting\_or\_structure & -0.2895 & 0.2619 & -1.106 & $0.269$ & -0.8028 & 0.2237 &  \\
 & harmful\_or\_illegal\_content & 0.8930 & 0.4147 & 2.153 & $0.031$ & 0.0802 & 1.7058 & \checkmark \\
 & rule\_or\_scope\_violations & -0.3686 & 0.1934 & -1.907 & $0.057$ & -0.7476 & 0.0103 &  \\
\bottomrule
\end{tabular}
\end{table*}

\begin{table*}[H]
\centering
\small
\caption{Main effects: Children (CH), Compliance. Linguistic PCs use FDR-adjusted $q$; moderator terms use nominal $p$.}
\label{tab:main_CH_compliance_prob}
\begin{tabular}{llrrrrrrc}
\toprule
\textbf{Type} & \textbf{Term} & $\hat{\beta}$ & \textbf{SE} & $z$ & $p$ & \textbf{95\% CI low} & \textbf{95\% CI high} & \textbf{Sig.} \\
\midrule
Intercept & Intercept & 1.7204 & 0.0417 & 41.280 & $<.001$ & 1.6387 & 1.8021 &  \\
\midrule
Moderator Type & modteam & -0.5688 & 0.0983 & -5.785 & $<.001$ & -0.7614 & -0.3761 & \checkmark \\
 & personalAccounts & -0.3896 & 0.0973 & -4.004 & $<.001$ & -0.5804 & -0.1989 & \checkmark \\
\midrule
Moderation Reason & content\_quality\_issues & 0.2227 & 0.0596 & 3.736 & $<.001$ & 0.1059 & 0.3395 & \checkmark \\
 & formatting\_or\_structure & 0.3162 & 0.0769 & 4.110 & $<.001$ & 0.1654 & 0.4669 & \checkmark \\
 & harmful\_or\_illegal\_content & -0.3404 & 0.2089 & -1.630 & $0.103$ & -0.7498 & 0.0690 &  \\
 & rule\_or\_scope\_violations & 0.1471 & 0.0720 & 2.044 & $0.041$ & 0.0060 & 0.2882 & \checkmark \\
\bottomrule
\end{tabular}
\end{table*}

\section*{B\quad Reference Category Effects (Account Issues)}

The following tables report the effects of each linguistic PC on the three
outcomes when moderation reason is the reference category (account issues).
All models use Bonferroni-corrected FDR at $q<.05$.

\begin{table*}[H]
\centering
\small
\caption{Account issues reference effects: High-Level (HL), Self-Censor.}
\label{tab:acct_HL_selfcensor_prob}
\begin{tabular}{lrrrrrrc}
\toprule
\textbf{PC} & $\hat{\beta}$ & \textbf{SE} & $z$ & $p$ & $p_\mathrm{FDR}$ & \textbf{95\% CI} & \textbf{Sig.} \\
\midrule
Conversational v.s Analytical & -0.0679 & 0.0101 & -6.716 & $<.001$ & $<.001$ & [-0.0877,\ -0.0481] & \checkmark \\
Sparse v.s Motivationally Rich & -0.0053 & 0.0084 & -0.636 & $0.525$ & $0.525$ & [-0.0217,\ 0.0111] &  \\
Detached Formal v.s Warm Social & 0.0292 & 0.0141 & 2.067 & $0.039$ & $0.063$ & [0.0015,\ 0.0569] &  \\
Vague Hedged v.s Concrete Specific & -0.0891 & 0.0143 & -6.245 & $<.001$ & $<.001$ & [-0.1171,\ -0.0611] & \checkmark \\
Elaborated v.s Quantified Direct & 0.0168 & 0.0129 & 1.306 & $0.192$ & $0.226$ & [-0.0084,\ 0.0420] &  \\
Cultural Authentic v.s Prohibitive Negation & 0.0694 & 0.0178 & 3.888 & $<.001$ & $<.001$ & [0.0344,\ 0.1043] & \checkmark \\
Formal Document v.s Numerical Authority & -0.0345 & 0.0237 & -1.458 & $0.145$ & $0.188$ & [-0.0809,\ 0.0119] &  \\
Brief v.s Long-form Elaborate & -0.0751 & 0.0251 & -2.993 & $0.003$ & $0.006$ & [-0.1243,\ -0.0259] & \checkmark \\
Lifestyle-Personal v.s Community-Scale Numerical & -0.1182 & 0.0240 & -4.934 & $<.001$ & $<.001$ & [-0.1652,\ -0.0712] & \checkmark \\
Cultural Formal v.s Netspeak Informal & 0.0506 & 0.0267 & 1.896 & $0.058$ & $0.084$ & [-0.0017,\ 0.1030] &  \\
Action-Oriented v.s Logically Elaborated & -0.0541 & 0.0210 & -2.575 & $0.010$ & $0.019$ & [-0.0953,\ -0.0129] & \checkmark \\
Prohibitive Negation v.s Descriptive Positive & 0.0995 & 0.0272 & 3.656 & $<.001$ & $<.001$ & [0.0462,\ 0.1529] & \checkmark \\
Perceptual-Structured v.s Adverbially Hedged & 0.0235 & 0.0326 & 0.720 & $0.471$ & $0.510$ & [-0.0404,\ 0.0874] &  \\
\bottomrule
\end{tabular}
\end{table*}

\begin{table*}[H]
\centering
\small
\caption{Account issues reference effects: High-Level (HL), Resistance.}
\label{tab:acct_HL_resistance_prob}
\begin{tabular}{lrrrrrrc}
\toprule
\textbf{PC} & $\hat{\beta}$ & \textbf{SE} & $z$ & $p$ & $p_\mathrm{FDR}$ & \textbf{95\% CI} & \textbf{Sig.} \\
\midrule
Conversational v.s Analytical & 0.2954 & 0.0414 & 7.131 & $<.001$ & $<.001$ & [0.2142,\ 0.3766] & \checkmark \\
Sparse v.s Motivationally Rich & -0.0010 & 0.0345 & -0.030 & $0.976$ & $0.976$ & [-0.0687,\ 0.0666] &  \\
Detached Formal v.s Warm Social & -0.1922 & 0.0595 & -3.232 & $0.001$ & $0.003$ & [-0.3088,\ -0.0757] & \checkmark \\
Vague Hedged v.s Concrete Specific & 0.1293 & 0.0621 & 2.080 & $0.037$ & $0.054$ & [0.0075,\ 0.2511] &  \\
Elaborated v.s Quantified Direct & -0.0097 & 0.0505 & -0.192 & $0.848$ & $0.919$ & [-0.1086,\ 0.0893] &  \\
Cultural Authentic v.s Prohibitive Negation & -0.1331 & 0.0628 & -2.120 & $0.034$ & $0.054$ & [-0.2562,\ -0.0101] &  \\
Formal Document v.s Numerical Authority & 0.2864 & 0.0868 & 3.300 & $0.001$ & $0.003$ & [0.1163,\ 0.4565] & \checkmark \\
Brief v.s Long-form Elaborate & 0.1850 & 0.0919 & 2.014 & $0.044$ & $0.057$ & [0.0049,\ 0.3651] &  \\
Lifestyle-Personal v.s Community-Scale Numerical & 0.3406 & 0.0952 & 3.578 & $<.001$ & $0.001$ & [0.1540,\ 0.5271] & \checkmark \\
Cultural Formal v.s Netspeak Informal & -0.2874 & 0.1159 & -2.480 & $0.013$ & $0.024$ & [-0.5146,\ -0.0603] & \checkmark \\
Action-Oriented v.s Logically Elaborated & 0.3360 & 0.0924 & 3.637 & $<.001$ & $0.001$ & [0.1549,\ 0.5170] & \checkmark \\
Prohibitive Negation v.s Descriptive Positive & -0.5436 & 0.1110 & -4.897 & $<.001$ & $<.001$ & [-0.7612,\ -0.3261] & \checkmark \\
Perceptual-Structured v.s Adverbially Hedged & -0.1390 & 0.1214 & -1.144 & $0.253$ & $0.298$ & [-0.3770,\ 0.0990] &  \\
\bottomrule
\end{tabular}
\end{table*}

\begin{table*}[H]
\centering
\small
\caption{Account issues reference effects: High-Level (HL), Compliance.}
\label{tab:acct_HL_compliance_prob}
\begin{tabular}{lrrrrrrc}
\toprule
\textbf{PC} & $\hat{\beta}$ & \textbf{SE} & $z$ & $p$ & $p_\mathrm{FDR}$ & \textbf{95\% CI} & \textbf{Sig.} \\
\midrule
Conversational v.s Analytical & -0.0750 & 0.0146 & -5.126 & $<.001$ & $<.001$ & [-0.1037,\ -0.0463] & \checkmark \\
Sparse v.s Motivationally Rich & 0.0179 & 0.0122 & 1.470 & $0.142$ & $0.204$ & [-0.0060,\ 0.0418] &  \\
Detached Formal v.s Warm Social & 0.0683 & 0.0212 & 3.217 & $0.001$ & $0.004$ & [0.0267,\ 0.1099] & \checkmark \\
Vague Hedged v.s Concrete Specific & 0.0175 & 0.0194 & 0.900 & $0.368$ & $0.479$ & [-0.0206,\ 0.0556] &  \\
Elaborated v.s Quantified Direct & 0.0103 & 0.0164 & 0.625 & $0.532$ & $0.629$ & [-0.0219,\ 0.0425] &  \\
Cultural Authentic v.s Prohibitive Negation & 0.0012 & 0.0218 & 0.053 & $0.958$ & $0.958$ & [-0.0416,\ 0.0439] &  \\
Formal Document v.s Numerical Authority & -0.1156 & 0.0321 & -3.606 & $<.001$ & $0.001$ & [-0.1784,\ -0.0528] & \checkmark \\
Brief v.s Long-form Elaborate & -0.0133 & 0.0283 & -0.471 & $0.637$ & $0.691$ & [-0.0687,\ 0.0421] &  \\
Lifestyle-Personal v.s Community-Scale Numerical & -0.0734 & 0.0360 & -2.036 & $0.042$ & $0.077$ & [-0.1440,\ -0.0028] &  \\
Cultural Formal v.s Netspeak Informal & 0.0836 & 0.0399 & 2.093 & $0.036$ & $0.077$ & [0.0053,\ 0.1618] &  \\
Action-Oriented v.s Logically Elaborated & -0.0922 & 0.0332 & -2.782 & $0.005$ & $0.014$ & [-0.1572,\ -0.0273] & \checkmark \\
Prohibitive Negation v.s Descriptive Positive & 0.1500 & 0.0383 & 3.912 & $<.001$ & $<.001$ & [0.0748,\ 0.2251] & \checkmark \\
Perceptual-Structured v.s Adverbially Hedged & 0.0756 & 0.0438 & 1.726 & $0.084$ & $0.137$ & [-0.0103,\ 0.1616] &  \\
\bottomrule
\end{tabular}
\end{table*}

\begin{table*}[H]
\centering
\small
\caption{Account issues reference effects: Children (CH), Self-Censor.}
\label{tab:acct_CH_selfcensor_prob}
\begin{tabular}{lrrrrrrc}
\toprule
\textbf{PC} & $\hat{\beta}$ & \textbf{SE} & $z$ & $p$ & $p_\mathrm{FDR}$ & \textbf{95\% CI} & \textbf{Sig.} \\
\midrule
Polite Reflective v.s Spatial Detached & 0.0243 & 0.0090 & 2.701 & $0.007$ & $0.016$ & [0.0067,\ 0.0419] & \checkmark \\
Impersonal v.s Direct Positive Address & -0.1028 & 0.0166 & -6.188 & $<.001$ & $<.001$ & [-0.1354,\ -0.0702] & \checkmark \\
Tech Direct v.s Tentatively Negative Paced & 0.0603 & 0.0196 & 3.082 & $0.002$ & $0.006$ & [0.0219,\ 0.0987] & \checkmark \\
Causal Authority v.s Work-Time Framing & -0.0381 & 0.0139 & -2.749 & $0.006$ & $0.016$ & [-0.0652,\ -0.0109] & \checkmark \\
Social Reference v.s Causal Explanation & 0.0798 & 0.0172 & 4.637 & $<.001$ & $<.001$ & [0.0461,\ 0.1135] & \checkmark \\
Visual Attentional v.s Contracted Need & 0.0884 & 0.0259 & 3.417 & $<.001$ & $0.002$ & [0.0377,\ 0.1391] & \checkmark \\
Obligatory Need v.s Achievement Positive & 0.0280 & 0.0325 & 0.862 & $0.389$ & $0.527$ & [-0.0357,\ 0.0917] &  \\
\bottomrule
\end{tabular}
\end{table*}

\begin{table*}[H]
\centering
\small
\caption{Account issues reference effects: Children (CH), Resistance.}
\label{tab:acct_CH_resistance_prob}
\begin{tabular}{lrrrrrrc}
\toprule
\textbf{PC} & $\hat{\beta}$ & \textbf{SE} & $z$ & $p$ & $p_\mathrm{FDR}$ & \textbf{95\% CI} & \textbf{Sig.} \\
\midrule
Polite Reflective v.s Spatial Detached & -0.0690 & 0.0321 & -2.150 & $0.032$ & $0.086$ & [-0.1320,\ -0.0061] &  \\
Impersonal v.s Direct Positive Address & 0.3116 & 0.0883 & 3.529 & $<.001$ & $0.005$ & [0.1386,\ 0.4847] & \checkmark \\
Tech Direct v.s Tentatively Negative Paced & -0.1186 & 0.0776 & -1.528 & $0.127$ & $0.228$ & [-0.2707,\ 0.0335] &  \\
Causal Authority v.s Work-Time Framing & 0.1932 & 0.0763 & 2.531 & $0.011$ & $0.036$ & [0.0436,\ 0.3427] & \checkmark \\
Social Reference v.s Causal Explanation & -0.2928 & 0.0850 & -3.444 & $<.001$ & $0.005$ & [-0.4595,\ -0.1262] & \checkmark \\
Visual Attentional v.s Contracted Need & -0.2746 & 0.1334 & -2.058 & $0.040$ & $0.094$ & [-0.5362,\ -0.0131] &  \\
Obligatory Need v.s Achievement Positive & -0.1916 & 0.1063 & -1.803 & $0.071$ & $0.151$ & [-0.4000,\ 0.0167] &  \\
\bottomrule
\end{tabular}
\end{table*}

\begin{table*}[H]
\centering
\small
\caption{Account issues reference effects: Children (CH), Compliance.}
\label{tab:acct_CH_compliance_prob}
\begin{tabular}{lrrrrrrc}
\toprule
\textbf{PC} & $\hat{\beta}$ & \textbf{SE} & $z$ & $p$ & $p_\mathrm{FDR}$ & \textbf{95\% CI} & \textbf{Sig.} \\
\midrule
Polite Reflective v.s Spatial Detached & 0.0200 & 0.0124 & 1.609 & $0.108$ & $0.204$ & [-0.0044,\ 0.0444] &  \\
Impersonal v.s Direct Positive Address & -0.0600 & 0.0262 & -2.288 & $0.022$ & $0.068$ & [-0.1115,\ -0.0086] &  \\
Tech Direct v.s Tentatively Negative Paced & -0.0056 & 0.0271 & -0.206 & $0.837$ & $0.837$ & [-0.0587,\ 0.0475] &  \\
Causal Authority v.s Work-Time Framing & -0.0566 & 0.0248 & -2.285 & $0.022$ & $0.068$ & [-0.1052,\ -0.0081] &  \\
Social Reference v.s Causal Explanation & 0.0901 & 0.0265 & 3.394 & $<.001$ & $0.006$ & [0.0381,\ 0.1421] & \checkmark \\
Visual Attentional v.s Contracted Need & 0.0451 & 0.0414 & 1.091 & $0.275$ & $0.444$ & [-0.0360,\ 0.1262] &  \\
Obligatory Need v.s Achievement Positive & 0.0745 & 0.0332 & 2.243 & $0.025$ & $0.068$ & [0.0094,\ 0.1397] &  \\
\bottomrule
\end{tabular}
\end{table*}

\section*{C\quad Effective Slopes by Moderation Reason}

Effective slopes represent the total effect of each linguistic PC on each
outcome within each moderation reason category (main effect $+$ interaction).
The reference category is account issues. Interaction $p$-values test whether
the slope differs significantly from the reference slope.

\begin{longtable}{llrrrr}
\caption{Effective slopes by moderation reason: High-Level (HL), Self-Censor. Reference = account issues (shown first per PC).} \label{tab:slopes_HL_selfcensor_prob} \\
\toprule
\textbf{PC} & \textbf{Mod.\ Reason} & \textbf{Eff.\ Slope} & \textbf{Interaction $\Delta$} & \textbf{Int.\ $p$} & \textbf{Main $p$} \\
\midrule
\endfirsthead
\toprule
\textbf{PC} & \textbf{Mod.\ Reason} & \textbf{Eff.\ Slope} & \textbf{Interaction $\Delta$} & \textbf{Int.\ $p$} & \textbf{Main $p$} \\
\midrule
\endhead
\midrule \multicolumn{6}{r}{\textit{Continued on next page}} \\
\endfoot
\bottomrule
\endlastfoot
Conversational v.s Analytical & Account Issues & -0.0679 & --- & --- & $<.001$ \\
 & Content Quality & -0.0361 & 0.0318 & $0.068$ & $<.001$ \\
 & Formatting/Structure & -0.0566 & 0.0113 & $0.450$ & $<.001$ \\
 & Harmful/Illegal & -0.0034 & 0.0645 & $0.004$ & $<.001$ \\
 & Rule/Scope Violations & -0.0567 & 0.0112 & $0.455$ & $<.001$ \\
 & Spam/Scams & -0.0681 & -0.0002 & $0.989$ & $<.001$ \\
\midrule
Sparse v.s Motivationally Rich & Account Issues & -0.0053 & --- & --- & $0.525$ \\
 & Content Quality & -0.0003 & 0.0051 & $0.847$ & $0.525$ \\
\end{longtable}

\begin{longtable}{llrrrr}
\caption{Effective slopes by moderation reason: High-Level (HL), Resistance. Reference = account issues (shown first per PC).} \label{tab:slopes_HL_resistance_prob} \\
\toprule
\textbf{PC} & \textbf{Mod.\ Reason} & \textbf{Eff.\ Slope} & \textbf{Interaction $\Delta$} & \textbf{Int.\ $p$} & \textbf{Main $p$} \\
\midrule
\endfirsthead
\toprule
\textbf{PC} & \textbf{Mod.\ Reason} & \textbf{Eff.\ Slope} & \textbf{Interaction $\Delta$} & \textbf{Int.\ $p$} & \textbf{Main $p$} \\
\midrule
\endhead
\midrule \multicolumn{6}{r}{\textit{Continued on next page}} \\
\endfoot
\bottomrule
\endlastfoot
Conversational v.s Analytical & Account Issues & 0.2954 & --- & --- & $<.001$ \\
 & Content Quality & 0.4954 & 0.2000 & $0.022$ & $<.001$ \\
 & Formatting/Structure & 0.3304 & 0.0350 & $0.730$ & $<.001$ \\
 & Harmful/Illegal & 0.3039 & 0.0085 & $0.930$ & $<.001$ \\
 & Rule/Scope Violations & 0.4032 & 0.1078 & $0.165$ & $<.001$ \\
 & Spam/Scams & 0.5064 & 0.2109 & $0.019$ & $<.001$ \\
\midrule
Sparse v.s Motivationally Rich & Account Issues & -0.0010 & --- & --- & $0.976$ \\
 & Content Quality & 0.1824 & 0.1835 & $0.062$ & $0.976$ \\
\end{longtable}

\begin{longtable}{llrrrr}
\caption{Effective slopes by moderation reason: High-Level (HL), Compliance. Reference = account issues (shown first per PC).} \label{tab:slopes_HL_compliance_prob} \\
\toprule
\textbf{PC} & \textbf{Mod.\ Reason} & \textbf{Eff.\ Slope} & \textbf{Interaction $\Delta$} & \textbf{Int.\ $p$} & \textbf{Main $p$} \\
\midrule
\endfirsthead
\toprule
\textbf{PC} & \textbf{Mod.\ Reason} & \textbf{Eff.\ Slope} & \textbf{Interaction $\Delta$} & \textbf{Int.\ $p$} & \textbf{Main $p$} \\
\midrule
\endhead
\midrule \multicolumn{6}{r}{\textit{Continued on next page}} \\
\endfoot
\bottomrule
\endlastfoot
Conversational v.s Analytical & Account Issues & -0.0750 & --- & --- & $<.001$ \\
 & Content Quality & -0.0598 & 0.0152 & $0.490$ & $<.001$ \\
 & Formatting/Structure & -0.0273 & 0.0477 & $0.078$ & $<.001$ \\
 & Harmful/Illegal & 0.0011 & 0.0761 & $0.020$ & $<.001$ \\
 & Rule/Scope Violations & -0.0397 & 0.0352 & $0.100$ & $<.001$ \\
 & Spam/Scams & -0.0695 & 0.0055 & $0.806$ & $<.001$ \\
\midrule
Sparse v.s Motivationally Rich & Account Issues & 0.0179 & --- & --- & $0.142$ \\
 & Content Quality & -0.0018 & -0.0197 & $0.456$ & $0.142$ \\
\end{longtable}

\begin{longtable}{llrrrr}
\caption{Effective slopes by moderation reason: Children (CH), Self-Censor. Reference = account issues (shown first per PC).} \label{tab:slopes_CH_selfcensor_prob} \\
\toprule
\textbf{PC} & \textbf{Mod.\ Reason} & \textbf{Eff.\ Slope} & \textbf{Interaction $\Delta$} & \textbf{Int.\ $p$} & \textbf{Main $p$} \\
\midrule
\endfirsthead
\toprule
\textbf{PC} & \textbf{Mod.\ Reason} & \textbf{Eff.\ Slope} & \textbf{Interaction $\Delta$} & \textbf{Int.\ $p$} & \textbf{Main $p$} \\
\midrule
\endhead
\midrule \multicolumn{6}{r}{\textit{Continued on next page}} \\
\endfoot
\bottomrule
\endlastfoot
Polite Reflective v.s Spatial Detached & Account Issues & 0.0243 & --- & --- & $0.007$ \\
 & Content Quality & 0.0244 & 0.0001 & $0.997$ & $0.007$ \\
 & Formatting/Structure & 0.0220 & -0.0023 & $0.885$ & $0.007$ \\
 & Harmful/Illegal & 0.0074 & -0.0169 & $0.672$ & $0.007$ \\
 & Rule/Scope Violations & 0.0239 & -0.0004 & $0.984$ & $0.007$ \\
 & Spam/Scams & 0.0488 & 0.0245 & $0.166$ & $0.007$ \\
\midrule
Impersonal v.s Direct Positive Address & Account Issues & -0.1028 & --- & --- & $<.001$ \\
 & Content Quality & -0.0861 & 0.0167 & $0.444$ & $<.001$ \\
\end{longtable}

\begin{longtable}{llrrrr}
\caption{Effective slopes by moderation reason: Children (CH), Resistance. Reference = account issues (shown first per PC).} \label{tab:slopes_CH_resistance_prob} \\
\toprule
\textbf{PC} & \textbf{Mod.\ Reason} & \textbf{Eff.\ Slope} & \textbf{Interaction $\Delta$} & \textbf{Int.\ $p$} & \textbf{Main $p$} \\
\midrule
\endfirsthead
\toprule
\textbf{PC} & \textbf{Mod.\ Reason} & \textbf{Eff.\ Slope} & \textbf{Interaction $\Delta$} & \textbf{Int.\ $p$} & \textbf{Main $p$} \\
\midrule
\endhead
\midrule \multicolumn{6}{r}{\textit{Continued on next page}} \\
\endfoot
\bottomrule
\endlastfoot
Polite Reflective v.s Spatial Detached & Account Issues & -0.0690 & --- & --- & $0.032$ \\
 & Content Quality & -0.0155 & 0.0536 & $0.467$ & $0.032$ \\
 & Formatting/Structure & -0.0946 & -0.0256 & $0.667$ & $0.032$ \\
 & Harmful/Illegal & 0.2000 & 0.2691 & $0.091$ & $0.032$ \\
 & Rule/Scope Violations & -0.0428 & 0.0262 & $0.669$ & $0.032$ \\
 & Spam/Scams & -0.0561 & 0.0130 & $0.846$ & $0.032$ \\
\midrule
Impersonal v.s Direct Positive Address & Account Issues & 0.3116 & --- & --- & $<.001$ \\
 & Content Quality & 0.6627 & 0.3510 & $0.008$ & $<.001$ \\
\end{longtable}

\begin{longtable}{llrrrr}
\caption{Effective slopes by moderation reason: Children (CH), Compliance. Reference = account issues (shown first per PC).} \label{tab:slopes_CH_compliance_prob} \\
\toprule
\textbf{PC} & \textbf{Mod.\ Reason} & \textbf{Eff.\ Slope} & \textbf{Interaction $\Delta$} & \textbf{Int.\ $p$} & \textbf{Main $p$} \\
\midrule
\endfirsthead
\toprule
\textbf{PC} & \textbf{Mod.\ Reason} & \textbf{Eff.\ Slope} & \textbf{Interaction $\Delta$} & \textbf{Int.\ $p$} & \textbf{Main $p$} \\
\midrule
\endhead
\midrule \multicolumn{6}{r}{\textit{Continued on next page}} \\
\endfoot
\bottomrule
\endlastfoot
Polite Reflective v.s Spatial Detached & Account Issues & 0.0200 & --- & --- & $0.108$ \\
 & Content Quality & -0.0592 & -0.0792 & $0.007$ & $0.108$ \\
 & Formatting/Structure & -0.0045 & -0.0246 & $0.206$ & $0.108$ \\
 & Harmful/Illegal & -0.1586 & -0.1786 & $0.055$ & $0.108$ \\
 & Rule/Scope Violations & -0.0247 & -0.0447 & $0.032$ & $0.108$ \\
 & Spam/Scams & -0.0335 & -0.0535 & $0.013$ & $0.108$ \\
\midrule
Impersonal v.s Direct Positive Address & Account Issues & -0.0600 & --- & --- & $0.022$ \\
 & Content Quality & -0.0574 & 0.0027 & $0.941$ & $0.022$ \\
\end{longtable}

\section*{D\quad FDR-Corrected Interaction Terms}

The following tables list all PC $\times$ moderation-reason interaction terms
from the logistic models, with Benjamini--Hochberg FDR correction applied
within each outcome $\times$ PCA-set block. $\checkmark$ denotes $q_\mathrm{FDR}<.05$.

\begin{longtable}{llrrrr}
\caption{FDR-corrected interactions: High-Level (HL), Self-Censor.} \label{tab:fdr_HL_selfcensor_prob} \\
\toprule
\textbf{PC} & \textbf{Mod.\ Reason} & $\hat{\beta}$ & $p$ & $q_\mathrm{FDR}$ & \textbf{Sig.} \\
\midrule
\endfirsthead
\toprule
\textbf{PC} & \textbf{Mod.\ Reason} & $\hat{\beta}$ & $p$ & $q_\mathrm{FDR}$ & \textbf{Sig.} \\
\midrule
\endhead
\midrule \multicolumn{6}{r}{\textit{Continued on next page}} \\
\endfoot
\bottomrule
\endlastfoot
Conversational v.s Analytical & Content Quality & 0.0318 & $0.068$ & $0.275$ &  \\
 & Formatting/Structure & 0.0113 & $0.450$ & $0.672$ &  \\
 & Harmful/Illegal & 0.0645 & $0.004$ & $0.110$ &  \\
 & Rule/Scope Violations & 0.0112 & $0.455$ & $0.672$ &  \\
 & Spam/Scams & -0.0002 & $0.989$ & $0.989$ &  \\
\midrule
Sparse v.s Motivationally Rich & Content Quality & 0.0051 & $0.847$ & $0.917$ &  \\
 & Formatting/Structure & 0.0193 & $0.297$ & $0.552$ &  \\
 & Harmful/Illegal & -0.0454 & $0.051$ & $0.254$ &  \\
\end{longtable}

\begin{longtable}{llrrrr}
\caption{FDR-corrected interactions: High-Level (HL), Resistance.} \label{tab:fdr_HL_resistance_prob} \\
\toprule
\textbf{PC} & \textbf{Mod.\ Reason} & $\hat{\beta}$ & $p$ & $q_\mathrm{FDR}$ & \textbf{Sig.} \\
\midrule
\endfirsthead
\toprule
\textbf{PC} & \textbf{Mod.\ Reason} & $\hat{\beta}$ & $p$ & $q_\mathrm{FDR}$ & \textbf{Sig.} \\
\midrule
\endhead
\midrule \multicolumn{6}{r}{\textit{Continued on next page}} \\
\endfoot
\bottomrule
\endlastfoot
Conversational v.s Analytical & Content Quality & 0.2000 & $0.022$ & $0.121$ &  \\
 & Formatting/Structure & 0.0350 & $0.730$ & $0.862$ &  \\
 & Harmful/Illegal & 0.0085 & $0.930$ & $0.960$ &  \\
 & Rule/Scope Violations & 0.1078 & $0.165$ & $0.345$ &  \\
 & Spam/Scams & 0.2109 & $0.019$ & $0.117$ &  \\
\midrule
Sparse v.s Motivationally Rich & Content Quality & 0.1835 & $0.062$ & $0.197$ &  \\
 & Formatting/Structure & -0.0471 & $0.593$ & $0.771$ &  \\
 & Harmful/Illegal & 0.1980 & $0.107$ & $0.280$ &  \\
\end{longtable}

\begin{longtable}{llrrrr}
\caption{FDR-corrected interactions: High-Level (HL), Compliance.} \label{tab:fdr_HL_compliance_prob} \\
\toprule
\textbf{PC} & \textbf{Mod.\ Reason} & $\hat{\beta}$ & $p$ & $q_\mathrm{FDR}$ & \textbf{Sig.} \\
\midrule
\endfirsthead
\toprule
\textbf{PC} & \textbf{Mod.\ Reason} & $\hat{\beta}$ & $p$ & $q_\mathrm{FDR}$ & \textbf{Sig.} \\
\midrule
\endhead
\midrule \multicolumn{6}{r}{\textit{Continued on next page}} \\
\endfoot
\bottomrule
\endlastfoot
Conversational v.s Analytical & Content Quality & 0.0152 & $0.490$ & $0.678$ &  \\
 & Formatting/Structure & 0.0477 & $0.078$ & $0.242$ &  \\
 & Harmful/Illegal & 0.0761 & $0.020$ & $0.140$ &  \\
 & Rule/Scope Violations & 0.0352 & $0.100$ & $0.272$ &  \\
 & Spam/Scams & 0.0055 & $0.806$ & $0.888$ &  \\
\midrule
Sparse v.s Motivationally Rich & Content Quality & -0.0197 & $0.456$ & $0.678$ &  \\
 & Formatting/Structure & 0.0017 & $0.951$ & $0.964$ &  \\
 & Harmful/Illegal & -0.1131 & $0.162$ & $0.340$ &  \\
\end{longtable}

\begin{longtable}{llrrrr}
\caption{FDR-corrected interactions: Children (CH), Self-Censor.} \label{tab:fdr_CH_selfcensor_prob} \\
\toprule
\textbf{PC} & \textbf{Mod.\ Reason} & $\hat{\beta}$ & $p$ & $q_\mathrm{FDR}$ & \textbf{Sig.} \\
\midrule
\endfirsthead
\toprule
\textbf{PC} & \textbf{Mod.\ Reason} & $\hat{\beta}$ & $p$ & $q_\mathrm{FDR}$ & \textbf{Sig.} \\
\midrule
\endhead
\midrule \multicolumn{6}{r}{\textit{Continued on next page}} \\
\endfoot
\bottomrule
\endlastfoot
Polite Reflective v.s Spatial Detached & Content Quality & 0.0001 & $0.997$ & $0.997$ &  \\
 & Formatting/Structure & -0.0023 & $0.885$ & $0.991$ &  \\
 & Harmful/Illegal & -0.0169 & $0.672$ & $0.841$ &  \\
 & Rule/Scope Violations & -0.0004 & $0.984$ & $0.997$ &  \\
 & Spam/Scams & 0.0245 & $0.166$ & $0.365$ &  \\
\midrule
Impersonal v.s Direct Positive Address & Content Quality & 0.0167 & $0.444$ & $0.714$ &  \\
 & Formatting/Structure & 0.0437 & $0.064$ & $0.202$ &  \\
 & Harmful/Illegal & 0.0846 & $0.004$ & $0.028$ & \checkmark \\
\end{longtable}

\begin{longtable}{llrrrr}
\caption{FDR-corrected interactions: Children (CH), Resistance.} \label{tab:fdr_CH_resistance_prob} \\
\toprule
\textbf{PC} & \textbf{Mod.\ Reason} & $\hat{\beta}$ & $p$ & $q_\mathrm{FDR}$ & \textbf{Sig.} \\
\midrule
\endfirsthead
\toprule
\textbf{PC} & \textbf{Mod.\ Reason} & $\hat{\beta}$ & $p$ & $q_\mathrm{FDR}$ & \textbf{Sig.} \\
\midrule
\endhead
\midrule \multicolumn{6}{r}{\textit{Continued on next page}} \\
\endfoot
\bottomrule
\endlastfoot
Polite Reflective v.s Spatial Detached & Content Quality & 0.0536 & $0.467$ & $0.662$ &  \\
 & Formatting/Structure & -0.0256 & $0.667$ & $0.820$ &  \\
 & Harmful/Illegal & 0.2691 & $0.091$ & $0.273$ &  \\
 & Rule/Scope Violations & 0.0262 & $0.669$ & $0.820$ &  \\
 & Spam/Scams & 0.0130 & $0.846$ & $0.917$ &  \\
\midrule
Impersonal v.s Direct Positive Address & Content Quality & 0.3510 & $0.008$ & $0.078$ &  \\
 & Formatting/Structure & 0.0185 & $0.922$ & $0.945$ &  \\
 & Harmful/Illegal & 0.3981 & $0.015$ & $0.106$ &  \\
\end{longtable}

\begin{longtable}{llrrrr}
\caption{FDR-corrected interactions: Children (CH), Compliance.} \label{tab:fdr_CH_compliance_prob} \\
\toprule
\textbf{PC} & \textbf{Mod.\ Reason} & $\hat{\beta}$ & $p$ & $q_\mathrm{FDR}$ & \textbf{Sig.} \\
\midrule
\endfirsthead
\toprule
\textbf{PC} & \textbf{Mod.\ Reason} & $\hat{\beta}$ & $p$ & $q_\mathrm{FDR}$ & \textbf{Sig.} \\
\midrule
\endhead
\midrule \multicolumn{6}{r}{\textit{Continued on next page}} \\
\endfoot
\bottomrule
\endlastfoot
Polite Reflective v.s Spatial Detached & Content Quality & -0.0792 & $0.007$ & $0.085$ &  \\
 & Formatting/Structure & -0.0246 & $0.206$ & $0.516$ &  \\
 & Harmful/Illegal & -0.1786 & $0.055$ & $0.231$ &  \\
 & Rule/Scope Violations & -0.0447 & $0.032$ & $0.165$ &  \\
 & Spam/Scams & -0.0535 & $0.013$ & $0.105$ &  \\
\midrule
Impersonal v.s Direct Positive Address & Content Quality & 0.0027 & $0.941$ & $0.944$ &  \\
 & Formatting/Structure & 0.0058 & $0.903$ & $0.942$ &  \\
 & Harmful/Illegal & -0.0360 & $0.498$ & $0.804$ &  \\
\end{longtable}
\end{document}